\documentclass[preprint]{aastex}
\usepackage{mkfig}

\begin{document}

\title{Observations of Disrupted CME Material Falling Back Into the Low Corona}

\author{Brian E. Wood\altaffilmark{1}, Jason E. Kooi\altaffilmark{2}}

\altaffiltext{1}{Naval Research Laboratory, Space Science Division,
  Washington, DC 20375, USA; brian.e.wood26.civ@us.navy.mil}
\altaffiltext{2}{Naval Research Laboratory, Remote Sensing Division,
  Washington, DC 20375, USA}


\begin{abstract}

     We present an empirical study of a disrupted CME, parts of
which fall back to the Sun, using observations from SOHO, STEREO-A,
and SDO.  At UT~18:00 on 2024~August~16, a slow CME is overtaken by a
faster CME.  A leg of the second CME carries part of the slower CME
out with it, resulting in an unusually well-defined flux rope leg
for this CME.  This second CME is observed in radio by the VLA, with
Faraday rotation measurements showing a clear
magnetic flux rope signature.  A strong response is also later
seen when the radio-observed line of sight enters the CME leg
enriched by material from the disrupted CME.  Outside this leg,
the rest of the disrupted CME simply disappears
and is replaced by a large number of small jet-like downflows.
We see clear evidence of this plasma falling back
to the low corona in EUV images from SDO, roughly 6--17 hours after
the CME is disrupted, with an inferred downward velocity of
$V=-30$ km~s$^{-1}$.  There is a clear temperature dependence, with
the downflows seen first in the 211~\AA\ bandpass, followed
successively by responses at 193~\AA, 171~\AA, and 304~\AA.
The downflows are much slower than would be expected for a
ballistic descent, so we model the downflows using a kinematic drag
model.  In the 304~\AA\ bandpass, coronal rain activity is triggered
by the downflowing CME material, suggesting that downflows from the
upper corona could be contributing to coronal rain
more generally.

\end{abstract}


\section{Introduction}

     The white light C2 and C3 coronagraphs that are part of
the Large Angle and Spectrometric Coronagraph (LASCO) instrument
on board the Solar and Heliospheric Observatory (SOHO) have
been monitoring the outer solar corona continuously since
1996 \citep{geb95}.  Similar coronagraphs on the
Solar Terrestrial Relations Observatory (STEREO) spacecraft
have provided additional coronal monitoring since STEREO's
2006 launch \citep{rah08}.  Movies of the
coronagraphic images show ubiquitous outflows at all scales,
representing the solar wind at early phases of its acceleration,
and transients embedded within it, ranging in size from
small jet-like blobs emanating from the tops of helmet
streamers \citep{ymw98,ymw00} to large coronal mass
ejections (CMEs) that can fill the entire field of view (FOV).

     Occasionally downflows are seen in coronagraphic
data, contrasting with the general sense of outflow.
The most commonly observed downflows are
of the sort first described by \citet{ymw99}, small
and faint $50-100$ km~s$^{-1}$ downflows concentrated around
the heliospheric current sheet (HCS), originating below a height
of 5.5~R$_{\odot}$ \citep{nrs02}.
Although some of these downflows may be associated with the
aftermath of recent CMEs, the downflows seem more generally
correlated with the Sun's large-scale field structure
\citep{nrs14}.  Although LASCO's C2 coronagraph
has provided the most extensive observations of these
events, they have also been studied using STEREO's COR1
coronagraph \citep{ph17}, and Parker Solar Probe (PSP)
has more recently studied them up close using the
Wide Field Imager on Solar Probe (WISPR) instrument
\citep{av25}.

     A natural interpretation of the narrow downflows is that
they are generated by small-scale reconnection, but reconnection
should yield flows in opposite directions from the reconnection
site, implying that there should be an upward-directed component
corresponding to the downflow.  In rare occasions
($\sim$2\% of cases) an outward flow is indeed observed above a
narrow downflow, forming a distinct class of ``in/out pairs''
\citep{nrs07}.  With streamer blobs, inflows, and in/out
pairs all clearly correlated with streamer belt structures, it is
often assumed that these are similar reconnection-related phenomena,
with height of origin and viewing geometry determining
the visibility of downflows and outflows generated by individual
small-scale reconnection events \citep{nrs09,esd17,ymw18,bjl20}.

     A class of larger coronal downflows observed in coronagraphic
images are the core fallback events described by \citet{ymw02},
which are basically cases where a trailing part of a CME
fails to escape the Sun and is observed to fall back toward it.
A more recent well-observed example is described by \citet{ncj13}.
This phenomenon may be worthy of more attention, as it
could have some relation to the broader and more popular topic of
confined eruptions, corresponding to large solar flares with no
associated CME, the idea being that the field topology in the low
corona somehow inhibits eruption \citep[e.g.,][]{lkh16,tl21,mdk23}.
Likewise, core fallback events
appear to be cases where the coronal field structure results in
at least a partial failure of the CME eruption, albeit at
unusually high heights.

     We here will be presenting an example of coronal downflows
associated with a disrupted CME from 2024~August~16, which does not
fit neatly into any of the aforementioned categories of downflow.
Rather than simply failing to escape the Sun like the core fallback
events, the slow CME in question is instead destroyed by a
faster, following CME.  Part of this ill-fated CME carried away with the
second CME and part of it simply disappears, replaced by a
blizzard of small-scale downflows, presumably indicating material
from the disrupted CME being funneled back down to the Sun by
small-scale reconnection processes.  Another
characteristic of this event that makes it worthy of study is
that we are actually able to perceive the downflow all
the way back in the low corona, using EUV images from
the Atmospheric Imaging Assembly (AIA) instrument on the Solar
Dynamics Observatory (SDO).  We are not aware of any previous
example of low corona EUV signatures of material from a
failed CME falling all the way back to the Sun from such a
large height.

     Finally, we will also be presenting radio Faraday
rotation (RFR) measurements made by the Karl G.\ Jansky
Very Large Array (VLA) of the National Radio Astronomy
Observatory (NRAO) during this solar activity.
One of the lines of sight being
monitored by the VLA happened to cross the CME responsible
for disrupting the slower CME.  We will present a complete
analysis of the RFR measurements of this CME, involving a
full 3-D reconstruction of the CME's magnetic flux rope
(MFR) structure based on available imaging data.

\section{Coronagraphic Observations}

     Our study of solar activity on 2024~August~16-17 will be
based on both white-light coronagraphic imaging of the upper
corona from SOHO/LASCO and STEREO-A, and EUV imaging of the
low corona from STEREO-A and SDO/AIA.  We focus first on the
coronagraphic data.  The primary LASCO coronagraph of interest
is the C2 coronagraph, covering plane-of-sky distances from
Sun-center of 1.5--6 R$_{\odot}$, but we will also utilize the
C3 coronagraph, covering 3.7--30 R$_{\odot}$ \citep{geb95}.
For STEREO-A, the primary coronagraph
of interest is the COR2-A coronagraph covering 2.5--15.6 R$_{\odot}$,
though COR1-A observations closer to the Sun from 1.4--4.0 R$_{\odot}$
will also be used \citep{rah08}.

\begin{figure}[t]
\plotfiddle{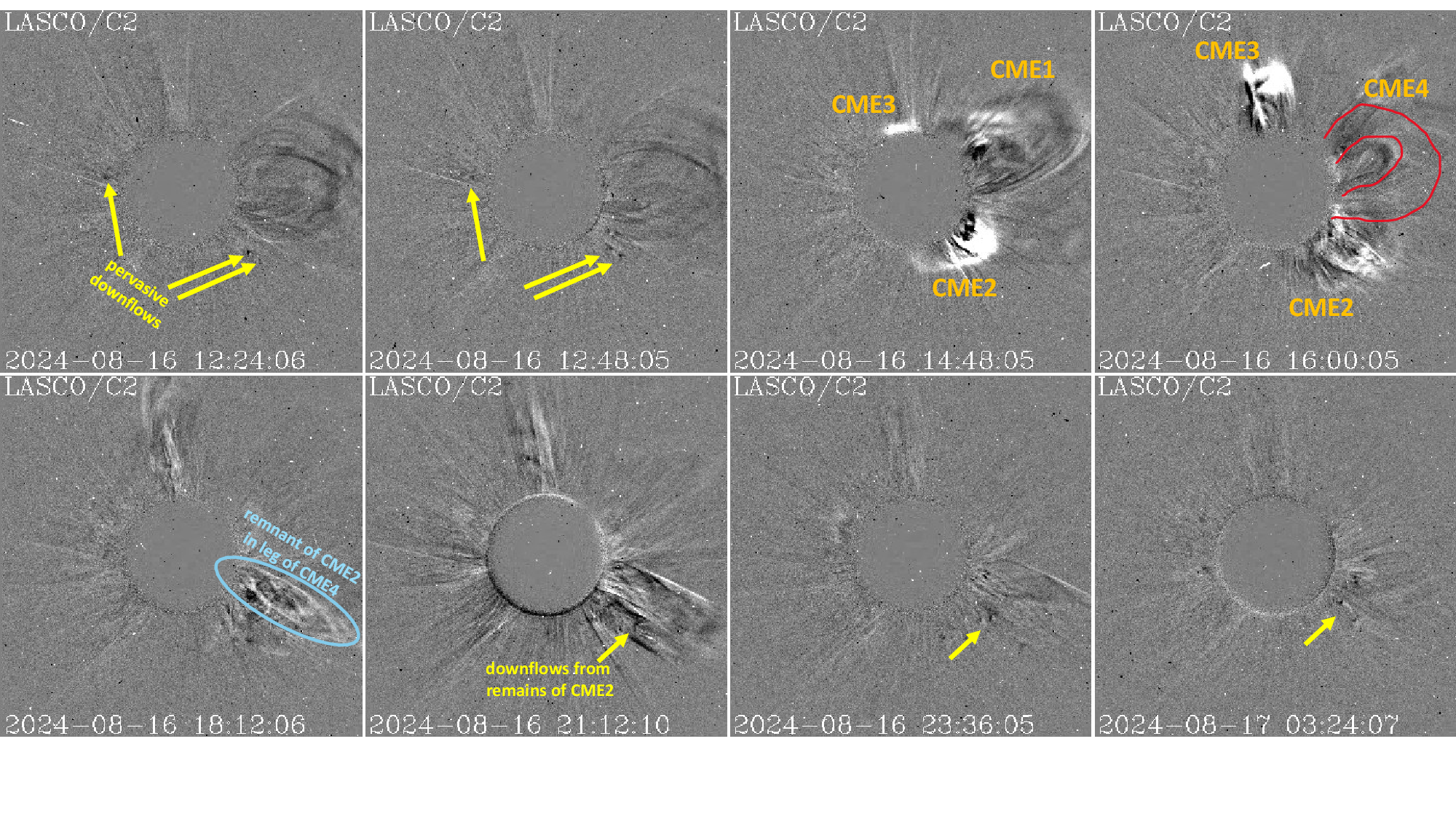}{2.4in}{0}{45}{45}{-215}{-35}
\caption{A sequence of LASCO/C2 images from 2024~August~16-17,
  focused on the eruption of four CMEs from the Sun, numbered
  CME1--4.  A rough outline (red) is provided for the faint CME4 in
  the fourth panel, assuming an MFR shape.  CME4 disrupts the
  slower CME2, with part of CME2 carried away by the southern
  leg of CME4 (fifth panel, blue ellipse), and with some of CME2's remains
  incorporated into numerous tiny downflows, one of which is
  indicated by yellow arrows in the last three panels.
  Abundant downflows are also present before the CMEs, as indicated
  in the first two panels.  A movie version of this figure is
  available in the online article, covering the time period
  from UT~4:48 on August~16 to UT~11:24 on August~17.}
\end{figure}
     Both SOHO/LASCO and STEREO-A observe the Sun from 1~au,
but while SOHO stays near Earth at the L1 Lagrangian point,
STEREO-A drifts in longitude relative to Earth by about
$22^{\circ}$ per year.  At the time of observation, STEREO-A was
relatively nearby, located only 22.6$^{\circ}$ ahead of Earth
in its orbit.  Figures~1-3 show sequences of SOHO and STEREO-A
coronagraphic images from 2024~August~16-17,
with Figures~1 and 2 displaying images from LASCO/C2 and LASCO/C3,
and Figure~3 showing one image from COR1-A and two from COR2-A.
Most of the images are displayed in a running difference format,
with the previous image subtracted from each image in order to
emphasize dynamic structures in the images.  The exception is
the single COR1-A image in Figure~3, where we instead use an
average-difference approach, subtracting a daily average COR1-A
image.

\begin{figure}[t]
\plotfiddle{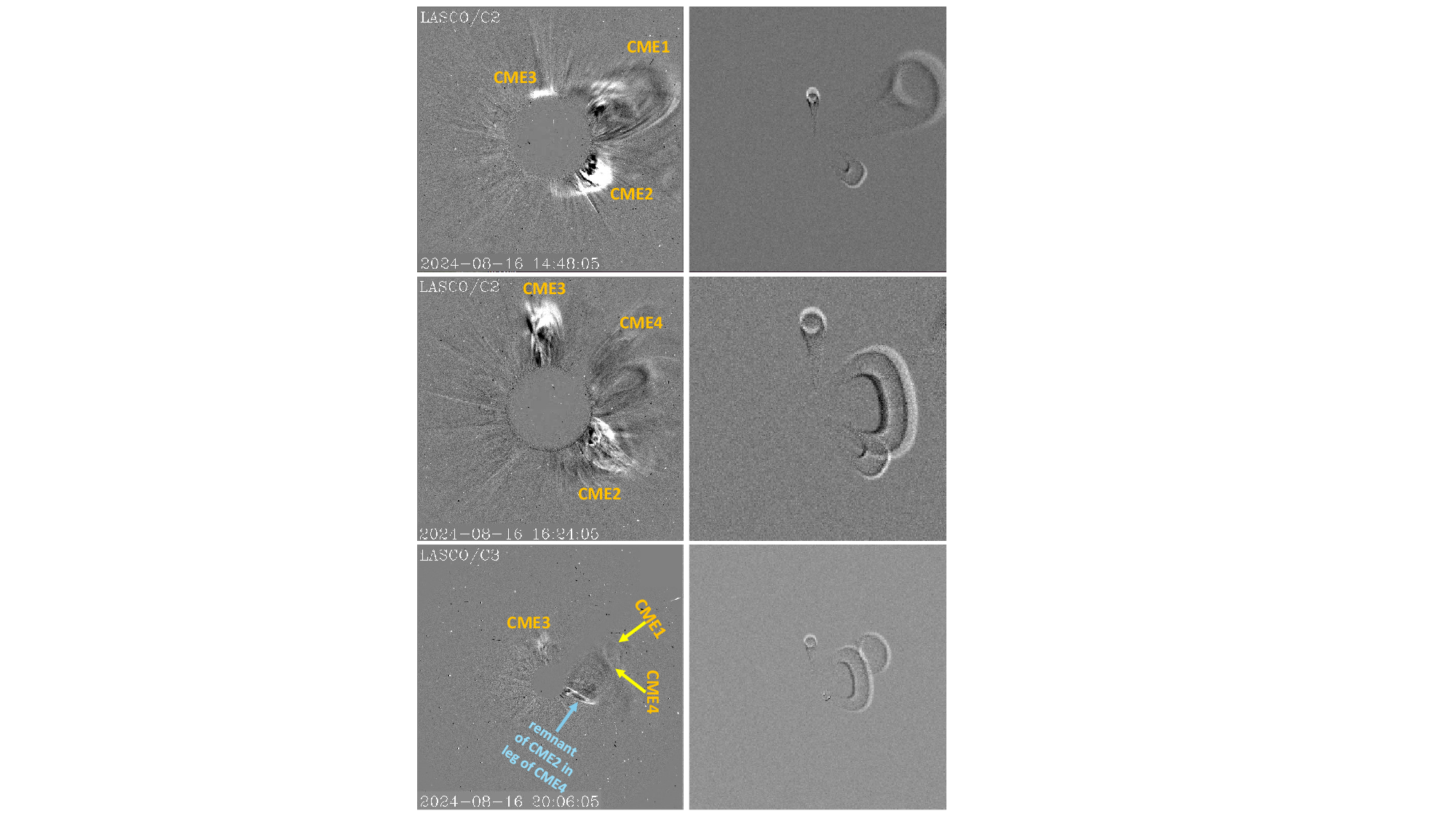}{3.2in}{0}{55}{55}{-245}{-15}
\caption{The left panels are three LASCO images of the
  four CMEs from 2024~August~16-17.  The right panels are
  synthetic images of these CMEs based on the 3-D kinematic and
  morphological reconstruction of the four CMEs described in
  Section~3.  The movie associated with Figure~6 provides a
  more complete comparison of the real and synthetic LASCO/C2
  images.}
\end{figure}
     The Figure~1 series of images is centered in time on a
sequence of 4 CMEs observed erupting from the Sun
between UT~13:30 and 15:00 on 2024~August~16.  We will refer to
these CMEs as CME1, CME2, CME3, and CME4; numbered in order of
appearance in LASCO C2.  Even before these CMEs, there are
unusually intense swarms of tiny downflows observed on both
east and west limbs of the Sun, of the sort described by 
\citet{ymw99} and more recently studied by PSP/WISPR \citep{av25}.
The first two panels of Figure~1
point to only a few of these, but the movie version of the
figure reveals that there are in fact dozens of them occurring
at all times.

     The primary CMEs that we will be studying are off
the west limb of the Sun.  These are not particularly bright
or dramatic eruptions, and in most respects they are typical
examples of CMEs that are routinely observed from an active Sun.
For example, the online Coordinated Data Analysis Workshops (CDAW)
CME catalog \citep[][https://cdaw.gsfc.nasa.gov/CME\_list]{sy04},
lists five west-limb CMEs in the 24 hours prior to our events,
between position angles $PA=225^{\circ}$ and $PA=315^{\circ}$.
Remnants of the last of these are visible in
the first panel of of Figure~1, representing the trailing parts
of a typical streamer blowout CME.

\begin{figure}[t]
\plotfiddle{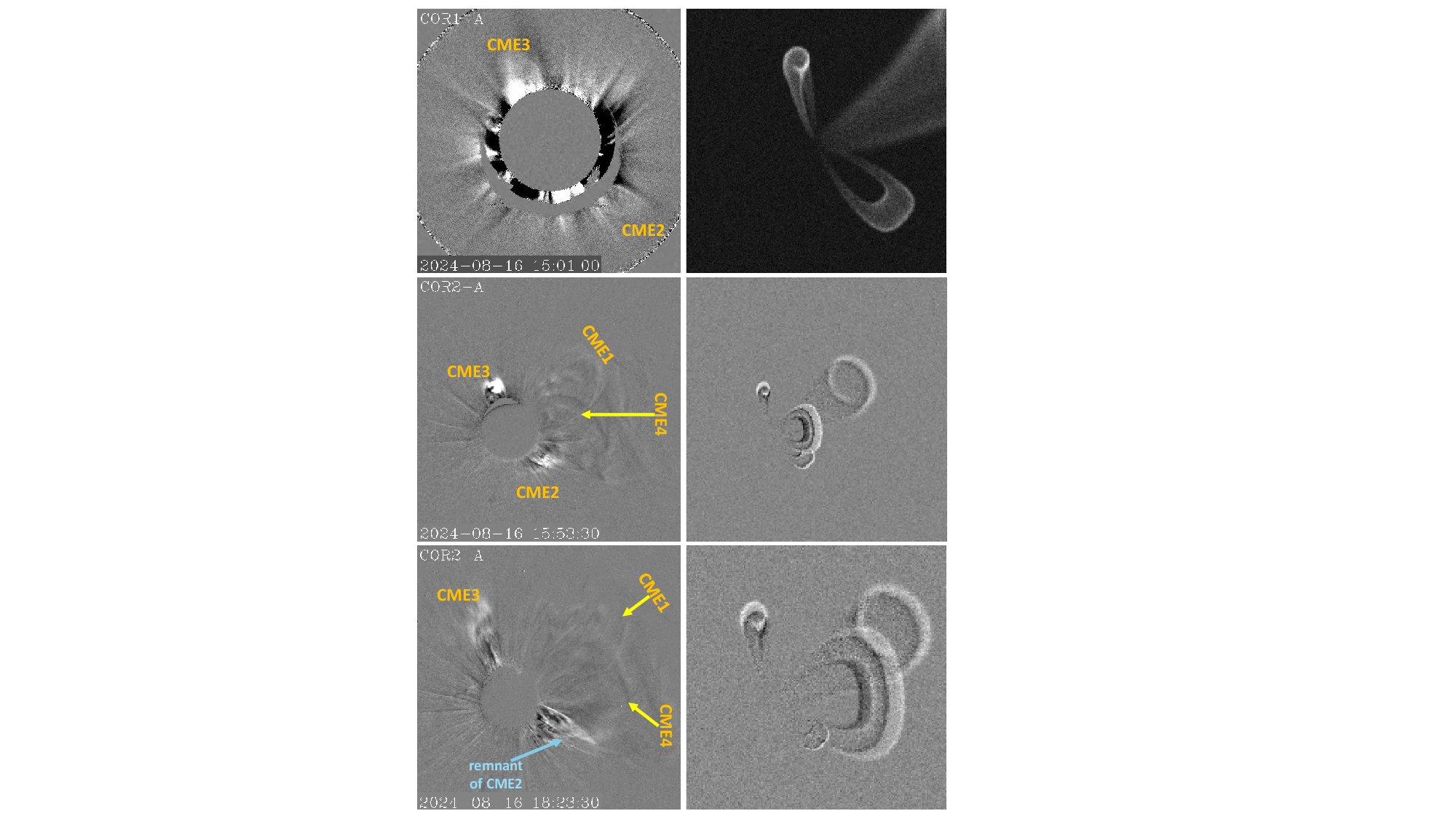}{3.2in}{0}{55}{55}{-245}{-15}
\caption{The left panels are three STEREO-A images of the
  four CMEs from 2024~August~16-17.  The right panels are
  synthetic images of these CMEs based on the 3-D kinematic and
  morphological reconstruction of the four CMEs described in
  Section~3.  The movie associated with Figure~6 provides a
  more complete comparison of the real and synthetic COR2-A
  images.}
\end{figure}
     The third panel of Figure~1 shows the first
three CMEs of interest.  The first one, CME1, erupts to the
west.  Slower CMEs are observed to the southwest (CME2) and to
the north (CME3).  The fourth and final CME of interest
(CME4) appears in the fourth panel of Figure~1, but despite
being relatively large it is nevertheless very faint and
hard to distinguish in still images, leading us to try to
outline it in red in the figure.  This outline indicates how we
are interpreting the CME's appearance as a N-S oriented
MFR, as MFR structures are commonly believed to lie at the
core of all CMEs \citep[see, e.g.,][]{rpl90,jc97,vb98,av13,bew17,bew25}.
Note that the limited field of view of LASCO/C2 makes
it difficult to discern CME4, which is more apparent in the
larger FOVs of LASCO/C3 (see Figure~2) and COR2-A (see Figure~3).

     Despite its faintness, it is actually CME4's interaction
with CME2 that is the focus of this article.  The southern
leg of CME4 appears to tear CME2 apart, with remnants of
the northern half of CME2 being carried outwards in the leg
of CME4.  This makes the southern leg of CME4 very bright (see
fifth panel of Figure~1 and third panel of Figure~2), in
contrast with the faintness of the rest of the CME.  Note that
all four of our CMEs are listed in the aforementioned CDAW
catalog, but the listing for CME4 only recognizes the bright southern
leg as being a distinct eruption.  Confusion about CME4 is
understandable.  Normally a CME is bright at its leading
edge, and its trailing parts are faint, but this
pattern is reversed for CME4.  Even though MFRs
are widely believed to exist at the core of all CMEs, clearly
identifying the legs of MFRs in coronagraph images is generally
very difficult.  Our CME4 provides an unusual opportunity to see
a CME flux rope leg very clearly, as this leg remains conspicuously
outlined in C2 images long after the leading edge of the CME has
left the C2 FOV.

     As for the seemingly obliterated CME2, in the third panel
of Figure~1 the CME seems to initially have a relatively bright
core with a front ahead of it, particularly visible extending
to the left of the core in the figure.  This would often be
interpreted as a shock or blast wave, but such a wave is
unexpected given that this is not a very fast CME.
(This will be shown explicitly in Section~3, where
kinematic measurements of all four CMEs will be presented.)
Regardless of structural interpretation, the entire CME2 structure
slows and disappears as the faster CME4 grazes it and
races past it.  As noted above, much of the northern half of
CME2 seems to be carried away in the southern leg of CME4, making
that leg very bright.  The southern half of CME2 simply
disappears, to be replaced by an intense blizzard of downflows,
not dissimilar in appearance from the swarm of downflows that
existed prior to the four CMEs.  We assume that these
downflows are of residual CME2 material falling back toward
the Sun, perhaps accelerated by small-scale reconnection back
toward the Sun.  The last 3 panels of Figure~1 point to one of
these tiny downflows, but only the movie version of the figure
properly demonstrates the pervasive nature of the downflows
after the disappearance of CME2.

\begin{figure}[t]
\plotfiddle{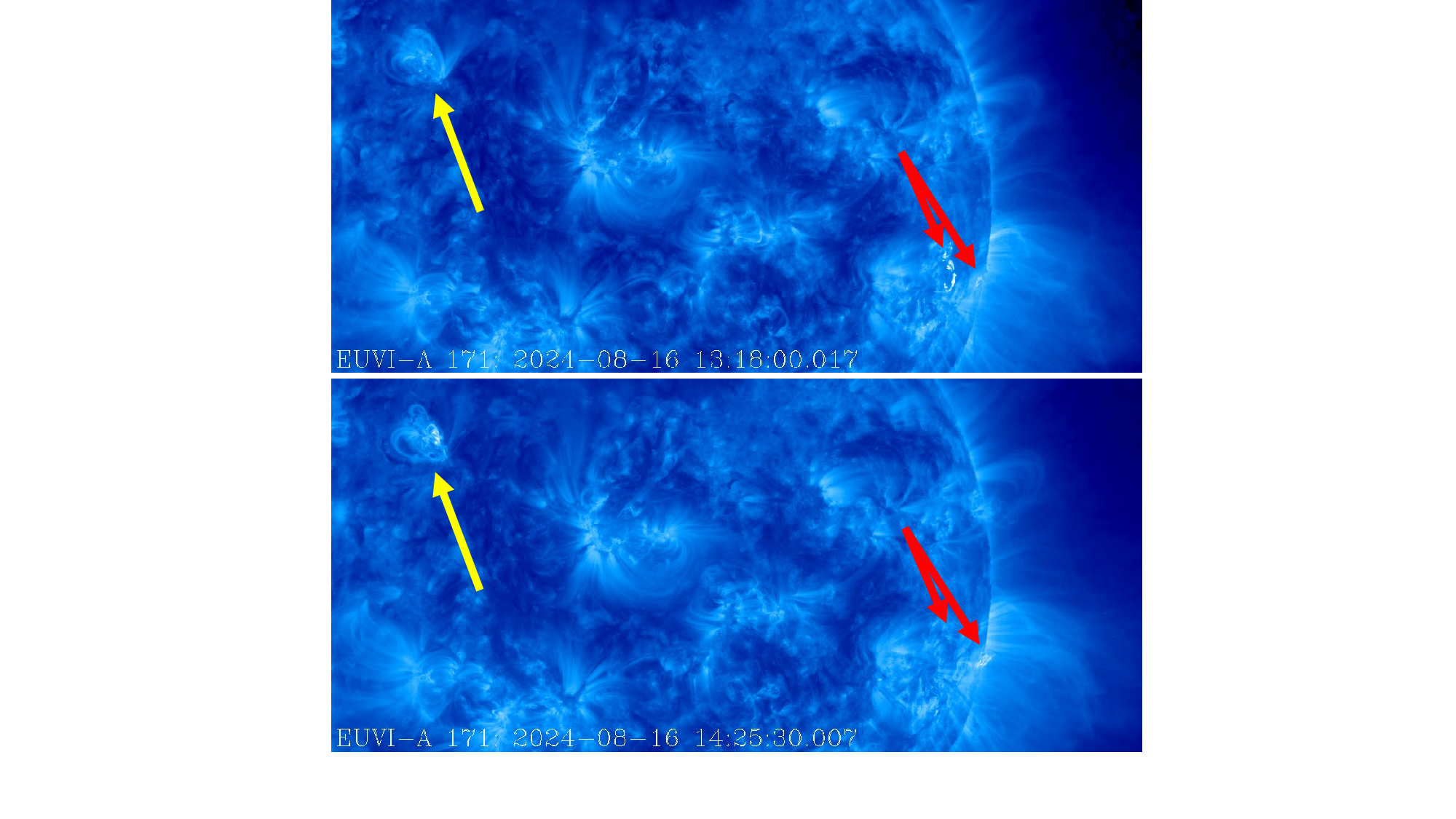}{3.2in}{0}{55}{55}{-265}{-35}
\caption{Two 171~\AA\ bandpass EUVI-A images from STEREO-A,
  with the red arrows pointing at surface activity associated
  with the initiation of CME1, CME2, and CME4.  The yellow
  arrow identifies the active region with activity associated
  with CME3.  A movie version of this figure is available in the
  online article, covering the time period
  UT~13:00-14:30 on August~16.}
\end{figure}
     We looked at EUV images of the Sun to try to
identify low corona signatures of the initiation of our
four CMEs, with limited success.  The source
region of CME1, CME2, and CME4 is near
the west limb of the Sun as viewed from Earth, so STEREO-A
at 22.6$^{\circ}$ west of Earth provides a better viewpoint
of this region than SDO.  Figure~4 shows two 171~\AA\ images
from STEREO-A's EUVI-A telescope \citep{rah08}.  Red
arrows point toward the location of brightenings and
prominence activity that occur from 13:00-14:30.  This is
the only clear candidate for activity associated with the
initiation of CME1, CME2, and CME4; with no way to
unambiguously determine which little brightenings
connect to which CME.  Only CME3 originates from a clearly
different location, identified with a yellow arrow in
Figure~4, where a filament eruption beginning at about
13:55 marks the inception of CME3.  We will return to
the subject of CME origins after the detailed kinematic
and morphological analysis of the next section.

\section{Kinematic and Morphological Reconstruction}

     Interpretation of how our four CMEs relate to and
interact with each other requires an analysis of their kinematics,
trajectories, and morphology.  The SOHO/LASCO and STEREO-A
data provide observations of the CMEs from two different
locations, yielding stereoscopic information about the
CMEs.  We use these data to perform a full 3-D kinematic
and morphological reconstruction of the events using techniques
used many times before \citep[e.g.,][]{bew09,bew17,bew25}.

     We start by discussing the kinematic measurements.  We
track the leading edge of each CME as it moves through the
FOV of each coronagraph.  For CME1, CME2, and CME3 we
use STEREO-A for these measurements, and consider both
COR1-A and COR2-A data.  For CME4 we instead use LASCO so
that we can take advantage of the particularly large C3 FOV.
(We might have chosen LASCO for CME1 for the same
reason, if the leading edge of CME1 was not partly obscured
by the occulter pylon in the C3 FOV.)  The kinematic
measurements initially are of elongation angle from
Sun-center ($\epsilon$) as a function of time.  It is
necessary to convert elongation angles to actual
Sun-center distances ($r$), which we do using the
``harmonic mean'' approximation of \citet{nl09},
\begin{equation}
r=\frac{2d\sin \epsilon}{1+\sin(\epsilon+\phi)},
\end{equation}
where $d$ is the distance from the observer to the Sun
and $\phi$ is the angle between the CME trajectory and the
observer's line-of-sight (LOS) to the Sun.  This equation
is derived assuming a CME can be approximated as a sphere
centered halfway between the Sun and the CME's leading edge.
The central trajectory and $\phi$ values for the two CMEs are
ultimately inferred from the morphological analysis that
will be described below.  The resulting distance measurements
versus time are plotted in Figure~5.

\begin{figure}[t]
\plotfiddle{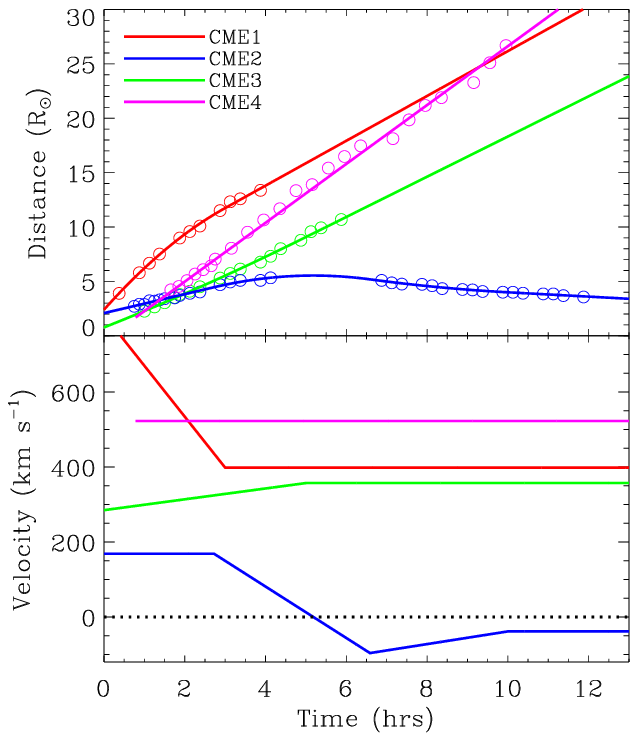}{3.0in}{0}{75}{75}{-225}{-285}
\caption{The top panel shows height versus time measurements
  for the four CMEs from 2024~August~16, based on STEREO-A
  coronagraphic measurements for CME1, CME2, and CME3;
  and SOHO/LASCO measurements for CME4.  The $t=0$ reference
  time is at UT~13:31.  CME4 is disrupted at about UT~18:00,
  and a downflow from its peak height is tracked afterward.
  A simple, multi-phase kinematic model
  is used to fit these data, yielding the velocity profiles
  for the CMEs in the bottom panel.}
\end{figure}
     In order to estimate leading edge velocities as a function
of time from these distance measurements, we assume a
simple multi-phase model where a CME's movement is approximated
as a phase of constant velocity or acceleration, ending in
a phase of constant velocity to allow for extrapolation to
times later than the observations.  We have in the past used
such kinematic models to track CMEs all the way to 1~au, if
desired \citep[e.g.,][]{bew17}.  The simplest such model
in Figure~5 is for CME4, where we see no evidence of any
acceleration and we simply assume a constant velocity, which
we measure to be $V=523$ km~s$^{-1}$.  For CME3, we see
evidence for modest acceleration, so the kinematic model
consists of a phase of constant acceleration followed
by a constant velocity phase, for which we infer a final
velocity of about $V=357$ km~s$^{-1}$.  CME1 has a more
unconventional kinematic profile compared with most CMEs
\citep[e.g.,][]{bew17}, as it seems to experience an
unusually strong deceleration in the COR2-A FOV, slowing
from about $V=700$ km~s$^{-1}$ down to a final speed
of about $V=398$ km~s$^{-1}$.

     Unsurprisingly, CME2 is the most awkward event to deal
with kinematically, due to the interaction with CME4 that
causes it to slow and then disappear, with some of its remnants
apparently falling back toward the Sun.  The kinematic model
of CME2 in Figure~5 has four phases, with an intial period of
constant velocity followed by one of constant decleration,
one of constant acceleration, and then a final one of constant
downflow velocity.  During the rising phase of CME2 we are
following the leading edge of the CME, while during the
falling phase we track a downflow that appears near where the
leading edge disappears.  It is dubious whether we are
actually tracking the same material, but the beginning and
ending velocities in Figure~5 represent estimates
of the average CME leading edge velocity in the COR2-A FOV,
which we estimate as $V=169$ km~s$^{-1}$, and the final speed
of downflows falling from the highest heights after the
CME disappears, which we estimate as $V=-38$ km~s$^{-1}$.
The kinematics of the downflowing material will be modeled
in more detail in Section~6.

\begin{figure}[t]
\plotfiddle{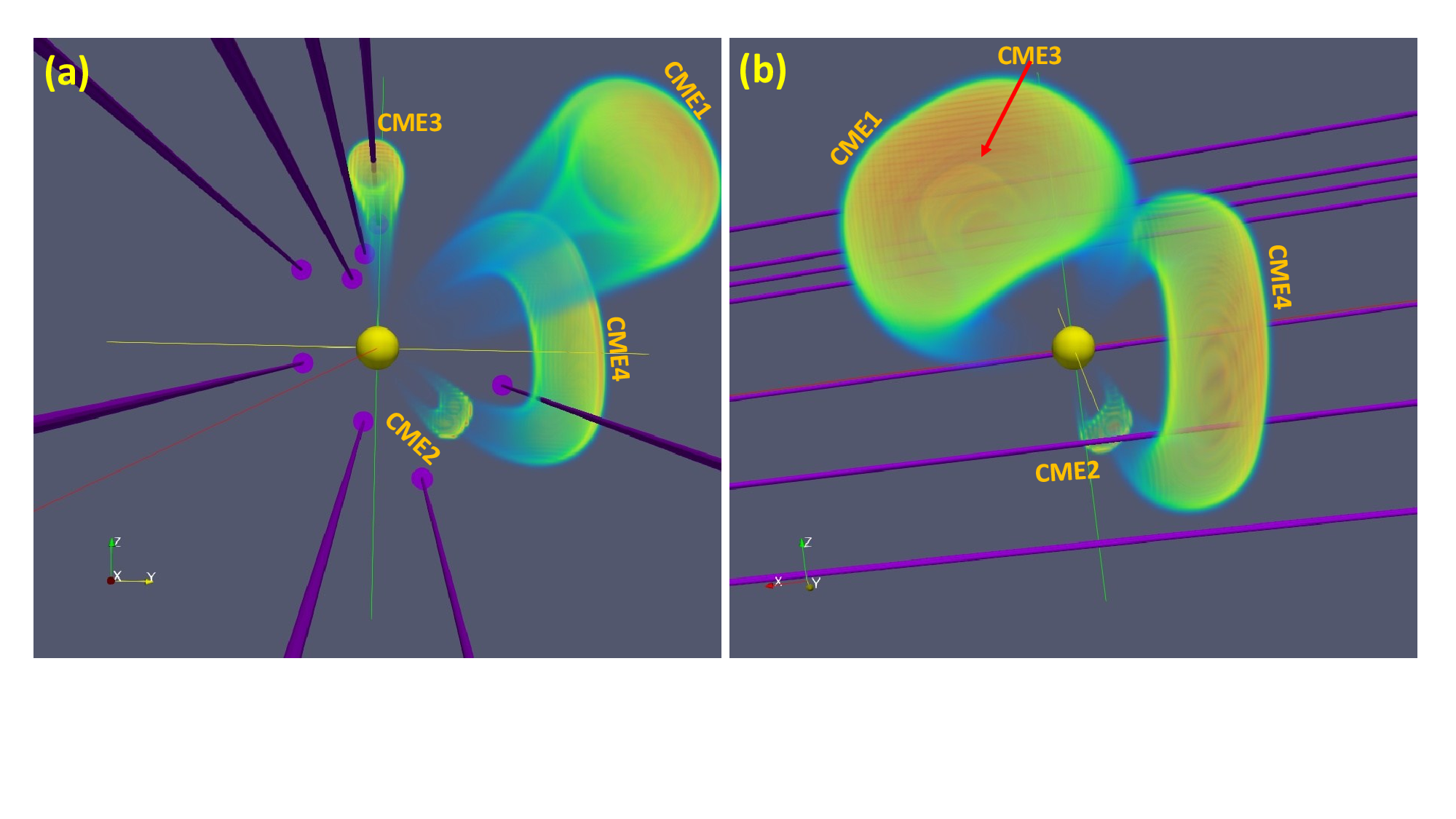}{2.5in}{0}{50}{50}{-240}{-65}
\caption{(a) A 3-D morphological reconstruction of the four
  CMEs erupting from the Sun on 2024~August~16, displayed for
  an HEE coordinate system with the x-axis pointing toward
  Earth and the z-axis toward ecliptic north.
  The relative positions of the CMEs are for a time of
  UT~18:53:30, after the disruption of CME2 by CME4.
  The size of the Sun is shown to scale.
  The viewpoint is approximately from Earth's perspective.
  Purple arrows indicate lines of sight to background
  radio sources monitored by the VLA at this time.  One of
  them intercepts CME3 and another intercepts CME4.
  (b) Same as (a), but for a viewpoint about $90^{\circ}$
  west of Earth.  An animation in the online version of this
  article provides both a more thorough presentation of the
  model CME eruptions, and also a thorough comparison of real and
  synthetic images for LASCO/C2 and COR2-A, covering
  the time period UT~13:36-22:23 on August~16.}
\end{figure}
     Turning from kinematics to morphology, we use the
stereoscopic imagery from SOHO/LASCO and STEREO-A to reconstruct
each CME's 3-D morphology, assuming an underlying MFR
shape.  We use a parametrized mathematical
prescription for generating 3-D MFR shapes that we have used
many times in the past.  This procedure is described in detail
by \citet{bew09}, but is most extensively utilized in
the STEREO CME survey of \citet{bew17}.  An assumed
set of MFR parameters yields a 3-D MFR shape.  This shape is
then used to generate a 3-D density cube, with mass placed only
on the surface of the MFR to outline the CME boundary.
We assume self-similar expansion for the MFR,
meaning that this single density cube applies at all times, with
only the axis scale changing in a manner described by the
kinematic model of the CME (e.g., Figure~5).  Thus, from the 3-D
density cube we can compute synthetic images of the CME for any FOV
from any perspective at any time, for comparison with the actual
images.  The synthetic images are computed using a white
light rendering routine that computes the Thomson scattering
within the density cube \citep{deb66,afrt06}.
Simple trial-and-error and subjective judgment are used to vary
the MFR parameters and decide which parameters yield synthetic images
that collectively best match the actual images.  Our final inferred
MFR morphologies are depicted in Figure~6, which shows two viewpoints
of the relative positions of the four CMEs at 18:53:30.
Synthetic images of the morphological reconstruction are shown in
Figures~2 and 3, for comparison with the actual images.  A movie
associated with Figure~6 provides a more detailed illustration of
the model CMEs erupting, along with a more thorough comparison of
real and synthetic LASCO/C2 and COR2-A images.

\begin{table}[t]
\small
\begin{center}
Table 1:  CME Morphological Parameters \\
\begin{tabular}{clcccc} \hline \hline
Parameter & Description & CME1 & CME2 & CME3 & CME4 \\
\hline
$\lambda_s$ (deg)& Trajectory longitude &  73   & 103  &   2  & 118  \\
$\beta_s$ (deg)  & Trajectory latitude  &  35   & -40  &  60  &   5  \\
$\gamma_s$ (deg) & Tilt angle           &  30   &  55  & -80  &  80  \\
FWHM$_s$ (deg)   & Angular width        & 66.3  & 46.1 & 47.4 & 94.2 \\
$\Lambda_s$      & Aspect ratio         & 0.16  & 0.13 & 0.11 & 0.12 \\ 
$\eta_s$         & Ellipticity          & 1.3   & 1.0  & 1.0  & 1.3  \\
$\alpha_s$       & Leading edge shape   &   3   &   4  &   3  &   6  \\
\hline
\end{tabular}
\end{center}
\end{table}
     Table~1 lists the MFR fit parameters for our four CMEs, using
the variable names from \citet{bew17}.  Briefly, $\lambda_s$ and
$\beta_s$ describe the central trajectories in
heliocentric-Earth-ecliptic (HEE) coordinates.  The $\gamma_s$
parameter indicates the tilt angle of the MFR, with
$\gamma_s=0^{\circ}$ corresponding to an E-W orientation parallel to
the ecliptic, and $\gamma_s>0^{\circ}$ indicating an upward tilt of
the west leg.  The FWHM$_s$ parameter is the
full-width-at-half-maximum angular width of the MFR.  The aspect
ratio, $\Lambda_s$, indicates the minor radius of the apex of the MFR
divided by the distance of the apex from the Sun, and so is a measure
of how fat the MFR is.  The ellipticity of the MFR channel is
described by $\eta_s$, which is the major radius divided by the minor
radius.  A value of $\eta_s=1$ would indicate a circular MFR cross
section.  Finally, the $\alpha_s$ parameter defines the shape of the
MFR leading edge, with higher values leading to flatter leading edges.

\begin{figure}[t]
\plotfiddle{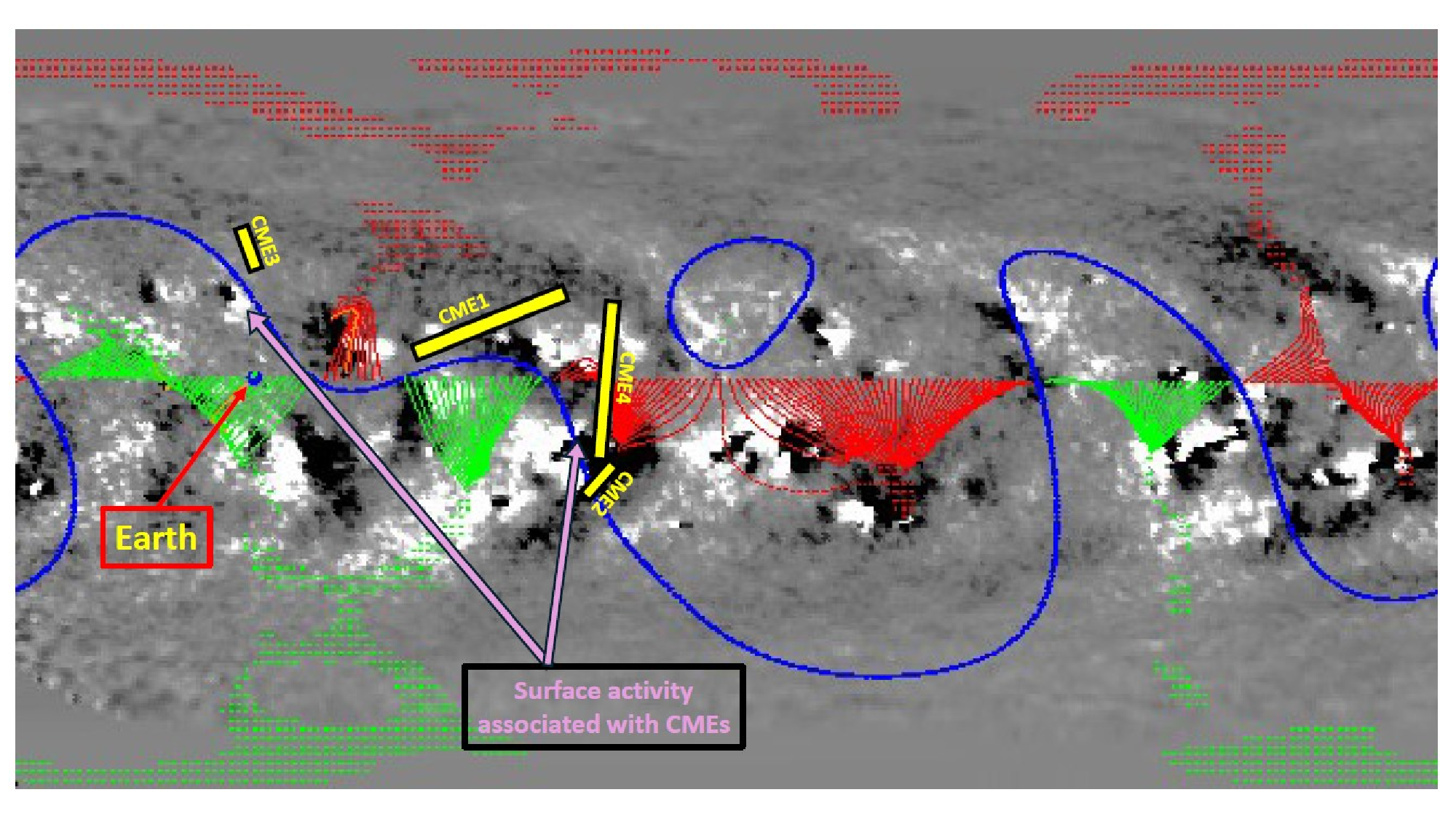}{2.8in}{0}{50}{50}{-240}{-20}
\caption{A 2024~August~16 full-disk solar magnetogram and a PFSS
  field model provided by the NSO Integrated Synoptic Program (NISP).
  The thick blue line indicates the HCS.  Green and red lines indicate
  how the ecliptic plane magnetically connects with the solar surface.
  The location of the Earth in the ecliptic is shown.
  Lavender arrows point to the two regions where we see
  surface activity that we associate with our four CMEs (see Fig.~4),
  which are directly under the HCS.  Yellow bars are used to
  roughly indicate the locations and orientations of CME1, CME2,
  CME3, and CME4, based on the reconstruction in Figure~6, implying
  a possible connection with the HCS.}
\end{figure}
     Given that they all erupt within a couple hours of each
other, the four CMEs we have analyzed here are likely related.
Figure~6 shows that CME1, CME2, and CME4 exhibit spatial
overlap.  The southern leg of CME4 intersects CME2, consistent
with this leg disrupting CME2 as CME4 erupts.
The northern leg of CME4 seems coincident with the northwestern
leg of CME1.

     In order to investigate this further, Figure~7 illustrates
how our CMEs relate to the HCS, as
estimated by a magnetogram and potential field source
surface (PFSS) model provided as a data product from the National
Solar Observatory (NSO) Integrated Synoptic Program (NISP)
(https://nso.edu/data/nisp-data).  The two active regions with the
activity that we associate with our CMEs (see Figure~4) appear
to be right along the HCS.  Our four CMEs also have trajectories
that are near the HCS, and they even have orientations that
roughly mirror it.  Of particular note is the $90^{\circ}$ turn
of the HCS from E-W to N-S that happens roughly $90^{\circ}$ west
of the Earth's location.  Our CME1 seems to be associated with the
E-W segment, and CME4 with the N-S part.  Connections
between CMEs and the HCS have been noted before, and can be
understood as being due to the HCS representing a minimum of
magnetic pressure, resulting in CMEs deflecting towards it \citep{ck13}.
A natural interpretation of these eruptions is
that they are symptoms of instability in this part of the HCS.
A connection among CME1, CME2, and CME4 is easy to believe, with
their spatial overlap and with related surface activity that
seemingly is associated with the same active region at the west
limb as viewed from Earth and STEREO-A.  Connecting CME3 with the
other CMEs is less certain, as it is farther away, and associated
with surface activity from a different active region, albeit
one also very near the HCS.

     A final point to make about Figure~7 is that it illustrates
how warped the HCS is at this time.  Relative to Earth's location,
there are places toward both the east and west limbs where the HCS
has a N-S orientation and is therefore viewed face-on.  This is
likely one factor behind the pervasiveness of downflows seen on both
limbs both before and after the CMEs erupt (see Figure~1).
A connection between downflows and HCS structure and orientation,
particularly a N-S oriented streamer belt, has been previously
noted \citep{nrs07,ymw18}.  This increase
in downflow visibility is in part because the face-on orientation
of the HCS spreads out the downflows across a wider range of
position angles, but it also increases the contrast with the
background.  When the HCS is roughly flat and oriented
more E-W, it will be viewed as a very bright streamer structure
in coronagraphic images, and it may be harder to pick out any
tiny downflows superposed on it, whereas if the HCS is viewed
face-on the background contrast is increased, allowing
such downflows to be more apparent.

\section{Triangulation of Downflows}

     In the previous section, we used stereoscopic
information from SOHO/LASCO and STEREO-A imaging to infer the
morphology and trajectory directions of the four 2024~August~16
CMEs that are the focus of our study.  Our methodology was
a forward modeling approach assuming a parameterized MFR shape
for the CME.  However, the little downflows that follow the
disappearance of CME2 offer an alternative approach.  Due to
their small size, their locations can be inferred by direct
triangulation rather than by forward modeling assuming some
parametrized shape.

\begin{figure}[t]
\plotfiddle{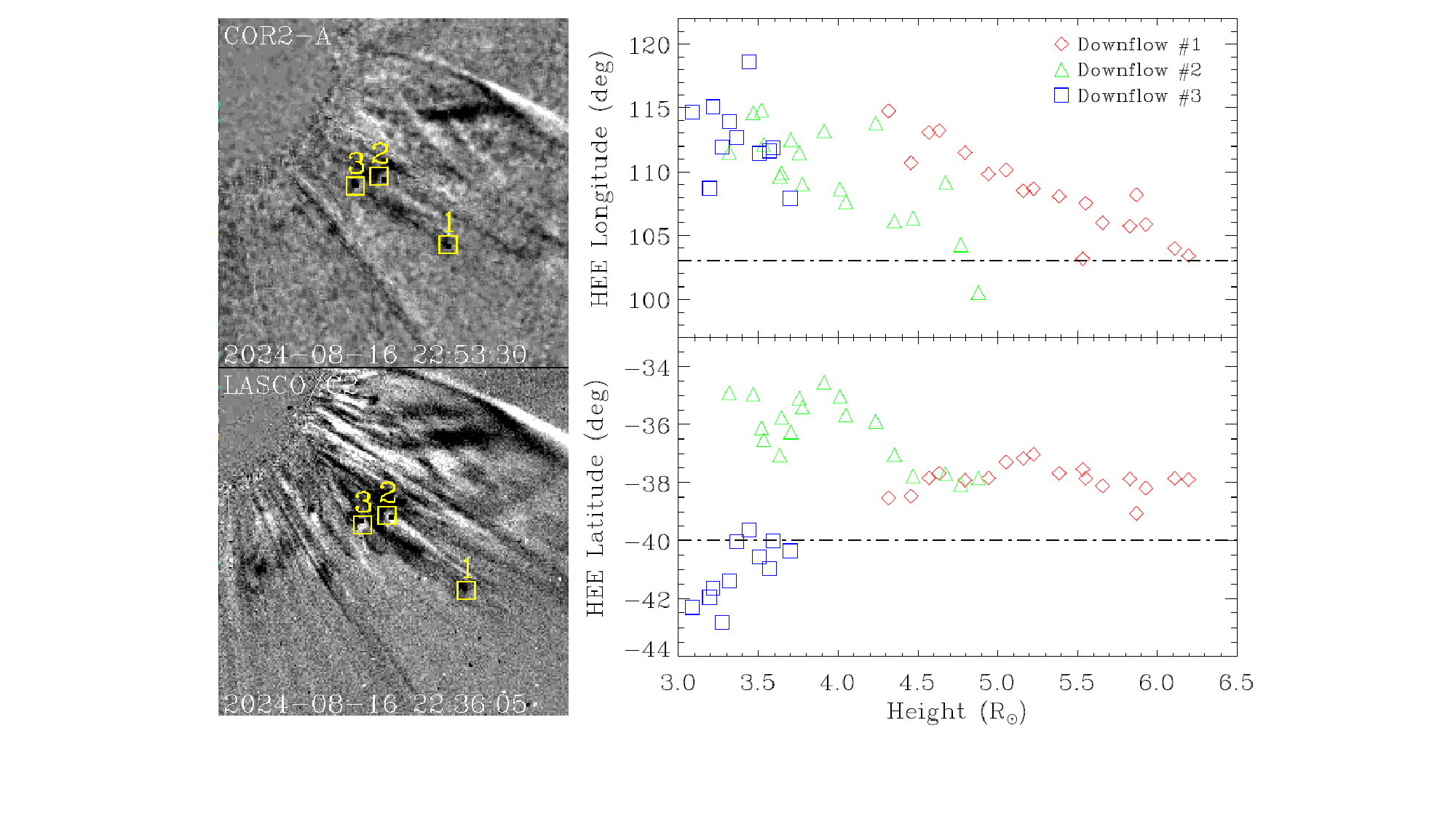}{3.4in}{0}{60}{60}{-280}{-45}
\caption{The left part of the figure shows a COR2-A and a
  LASCO/C2 image from similar times, with boxes outlining
  three small downflows that are triangulated and tracked
  with time, numbered \#1-\#3.  The right side of the
  figure plots the longitude and latitude of the three
  downflows versus height, in HEE coordinates.  The dot-dashed
  lines indicate the central trajectory longitude and latitude
  of CME2, the disruption of which leads to the downflows.}
\end{figure}
     In this section, we identify three particularly prominent
downflows that follow the disruption of CME2, ones that are
trackable for at least a few hours.  These are shown explicitly
in Figure~8, with the little downflow features outlined in
both a COR2-A image and a LASCO/C2 image from close to the
same time.  The relatively small SOHO/STEREO-A longitudinal
separation of $22.6^{\circ}$ is less than ideal for the kind
of forward modeling performed in the last section, but is
advantageous for triangulation techniques because the similarity
in viewpoint increases confidence in connecting tiny features
in COR2-A with the corresponding features in LASCO/C2.
Measuring image positions of these features for COR2-A/C2 image
pairs allows for a triangulation of 3-D position as a function
of time.  This calculation is analogous to ``tiepointing''
analyses of erupting filamentary structures that have been
done in the past \citep{pcl09,bew16}.

     The right half of Figure~8 plots the inferred longitude
and latitude of the three downflows versus height, in HEE
coordinates.  The coordinates of all three are close to the
inferred central trajectory of CME2 listed in Table~1,
$(\lambda_s,\beta_s)=(103,-40)$.  Aside from supporting our
connection between the flurry of downflows seen after CME2
disappears with CME2 remnant material being funneled back
down toward the Sun, it also shows consistency
between the forward modeling CME morphological analysis, and
the more direct downflow triangulation measurements.

     There is some evidence that the flows are not purely
radial, with longitude and/or latitude changing as the
material moves downward.  For example, the longitude of
downflow \#1 appears to shift from $\lambda\approx 105^{\circ}$
to $\lambda\approx 115^{\circ}$ between $R=6.0$~R$_{\odot}$ and
$R=4.5$~R$_{\odot}$.  This could be indicating a nonradial
orientation of the field lines along which the flows are
funneling back down to the Sun.

\section{Radio Observations of CME4}

     A primary reason that our attention was drawn to the eruptive
activity occurring on 2024~August~16 was that we had
obtained time on the VLA to monitor the solar corona in this
period (project code VLA/24A-135).  Between UT~16:00 and UT~22:00,
the VLA was used to monitor eight bright
linearly polarized radio sources around the Sun, as
shown in Figure~6, in hopes that one or more of these VLA LOS's
would encounter a CME.  The VLA cycled through these eight separate
pointings roughly once every 30 minutes.  Radio observations,
specifically RFR measurements, represent one of the only means by
which magnetic field strength and orientation can be probed
remotely from Earth.  Thus, RFR observations of CMEs offer a
unique opportunity to study CME field structure, and test the
prevailing MFR paradigm for CMEs \citep{mkb85,jek17,eaj18,bew20b,jek22,eaj25}.

     Figure~6 shows that two of the radio LOS's do indeed sample
two of our four CMEs.  One LOS toward a radio galaxy at J2000
coordinates (RA,DEC)=(09:37:33.36,+13:31:00.4) encounters CME4,
and another LOS toward (RA,DEC)=(09:49:49.11,+15:22:03.2)
encounters CME3.  We will be focusing only on the CME4
observations in this particular paper, with its primary
focus on CME2 and CME4.  The CME3 study will be reserved for
a radio-focused article.

     The RFR diagnostic involves the detection of a change in
polarization position angle ($\chi$; defined by the electric field
vector) induced by the passage of a CME in front of the background
source.  Numerically, the rotation is
\begin{equation}
\Delta\chi=\left[ \left(\frac{e^3}{2\pi m_e^2 c^4} \right)
  \int_{LOS} n_e {\bf B\cdot ds} \right]\lambda^2 = [RM]\lambda^2,
\end{equation}
where $\lambda$ is the observed radio wavelength, and ${\bf ds}$
is the differential direction vector along the LOS.  The term
in square brackets is the rotation measure ($RM$), with
units of rad~m$^{-2}$.  The constant within the parentheses
includes the electron charge ($e$), electron mass ($m_e$), and
speed of light ($c$).  It is $RM$ that is the quantity
of interest, which diagnoses the integration of the density times
the parallel component of field along the LOS through the CME.
Thus, $RM$ depends on both the strength and orientation of
the field within the CME, as well as on its internal electron
density.

\begin{figure}[t]
\plotfiddle{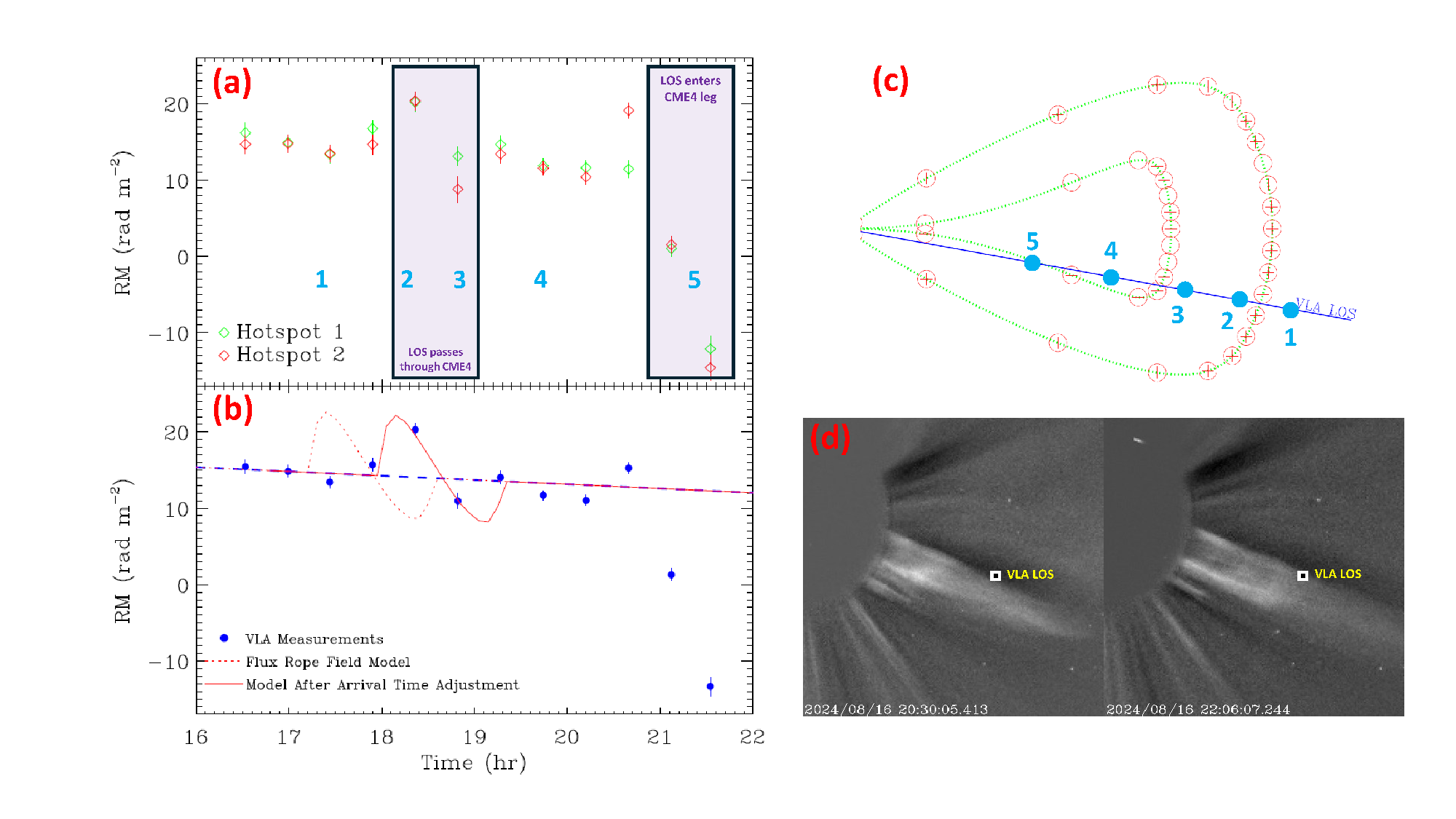}{2.8in}{0}{50}{50}{-250}{-30}
\caption{(a) Radio rotation measures versus time measured by VLA
  for an LOS through CME4.  The background radio galaxy has two
  bright lobes separated by $46.7^{\prime\prime}$, so separate $RM$ values
  are plotted for both hotspots.  As indicated by the blue numbers,
  which connect with the schematic CME4 depiction in panel (c), two
  separate signature of CME4
  are identified, a positive-negative signal indicative of the top
  of the CME4 MFR passing over the LOS, and at the end a very
  negative RM signal as the LOS grazes the bright southern leg of
  the MFR.  (b) An average $RM$ of the two hot spots is plotted, and
  compared with predictions from a model MFR (see Figures 6 and 10).
  An arbitrary 45-minute shift is necessary to get the arrival time
  correct.  (c) A schematic illustration of how the radio LOS tracks
  through CME4, first through the top of the MFR and later entering the
  southern leg.  Blue numbers connect with radio observing times in
  panel (a).  (d) Two LASCO/C3 images of the leg of CME4, showing
  the VLA LOS just before and after it enters the leg.}
\end{figure}
     The background radio galaxy
that VLA is using to sample CME4 consists of two bright lobes
separated by $46.7^{\prime\prime}$.  Figure~9(a) plots separate $RM$
measurements observed toward these two hotspots.  Guided by the
LASCO images and our 3-D reconstruction of CME4, we identify two
separate signatures of CME4 in the $RM$ measurements.  The
first is a positive $RM$ shift relative to the background followed
immediately by a negative shift, indicative of the LOS passing
through an MFR oriented perpendicular to the LOS, as suggested for
CME4 by Figure~6.  The sign change in $RM$ provides strong support
for the MFR model for this CME, as in the MFR paradigm this shift
in sign is naturally explained by the wrapping
of the field around the central axis of the MFR.  A positive $RM$
signature (beginning near UT~17:54) indicates that this azimuthal
field is pointed toward Earth at the CME4 leading edge, with the
inner edge of the MFR therefore having field pointed away from Earth,
leading to the following negature $RM$ shift at UT~18:49.
Figure~9(c) provides a schematic picture of this.
The $\pm 6$ rad~m$^{-2}$ variation in RM consistent with the initial
passage of CME4 is likely not due to residual ionospheric RFR.  The
ionospheric RFR is removed as part of the radio data calibration
procedure and, therefore, the total contribution is expected to
be negligible ($<0.1$ rad~m$^{-2}$) \citep{jek17}.

     After the positive-negative MFR signature, the LOS leaves CME4,
but a second CME4 signature is later observed when the LOS grazes
the bright southern leg of CME4, which is bright because it contains
the dense remnants of CME2.  Figure~9(d) uses two LASCO C3 images
to explicitly show the VLA LOS just before and after it enters
the CME leg at the end of the VLA observing period.  The result is
a large negative shift in $RM$.  In the MFR picture, the azimuthal
field at the inner edge should be in the same direction along the
entire inner edge of the MFR.  This is consistent with the negative
$RM$ shifts observed both after passage through the top of the MFR
and later along the inner part of the southern leg.  The more
dramatic signature in the leg is simply due to the leg being much
denser than the top of the CME, thanks to the remnants of CME2.
The negative $RM$ shift is roughly a factor of 4 larger in the
leg, suggesting densities about a factor of 4 higher there
than at the back of the top of the MFR.

     It is worth noting that the modest $RM$ differences
between the two hotspots may be indicative of real fine structure in
the field and density distribution within the corona and CME4.  For
example, there is significant discrepancy between the two as the LOS is
nearing the CME leg.  Nevertheless, the two independent measurements
are generally consistent, and in Figure~9(b) mean $RM$ values of
the two hotspots are shown.  It is these values that we use to
compare with the predictions of MFR models that we now compute
based on the 3-D CME4 reconstruction described in Section~3.
Our field insertion prescription is based on one we
have used for a number of previous studies \citep{bew20b,bew25},
which traces its roots to the physical MFR
model of \citet{tnc18}.  We refer the reader
to \citet{bew20b} in particular for details.

\begin{figure}[t]
\plotfiddle{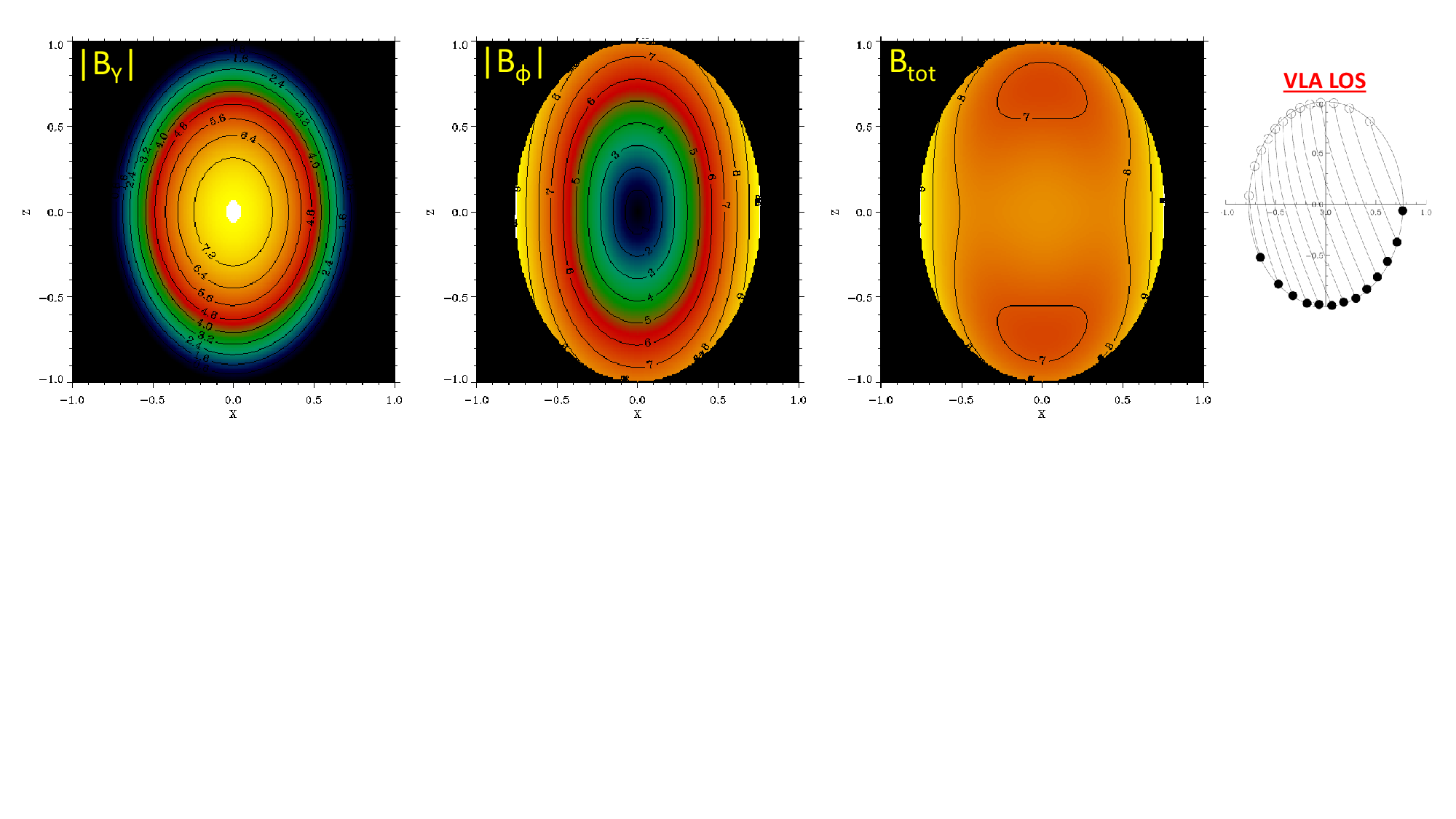}{1.6in}{0}{50}{50}{-245}{-135}
\caption{Maps of the magnetic field at the apex of the MFR for CME4,
  based on a model that fits the VLA $RM$ measurements (see Figure~9).
  From left to right, the maps show the axial field ($|B_Y|$), the
  azimuthal field along the elliptical contours of the MFR channel
  ($|B_{\phi}|$), and the total field ($B_{tot}$).  The contours
  indicate the field values in units of nT.  The values shown
  correspond to a time when the leading edge of the CME would
  reach 1~au.  The rightmost panel shows how the VLA LOS through
  the MFR channel changes with time, moving from right to left
  in 6-minute increments, with the VLA LOS from Earth traversing
  the MFR channel from the open circles to the closed circles.}
\end{figure}
     With the morphology and kinematics of the CME4 MFR
predetermined by the image-based reconstruction, there are
actually only two free parameters required to insert a plausible
field structure into the MFR.  The first is the axial field at
the MFR center, $B_t$, and the maximum azimuthal field at the
surface of the MFR, $B_p$.  Both of these parameters are defined
at the apex of the MFR, and are used to map the field at the apex,
using the \citet{tnc18} model.  Figure~10 shows
maps of axial field, $B_Y$, and azimuthal field $B_{\phi}$,
based on $B_t=+8$~nT and $B_p=+10$~nT, values that we will show
below can reproduce the observed $RM$ signature.
The axial field has the peak value of $B_t$ at MFR center, decreasing
to zero at the MFR surface.  In contrast, $B_{\phi}$ is zero
at the axis but increases toward the MFR surface, with a maximum,
$B_p$, at the surface along the minor axis of the ellipse.  As a
sign convention, we assume the positive $B_t$ corresponds to a
direction out of the plane for $B_Y$, and a positive $B_p$ corresponds
to a right-handed MFR, meaning the $B_{\phi}$ field lines follow
elliptical contours in a counter-clockwise direction in Figure~10.

     From the apex field maps of Figure~10, the field can be defined
throughout the 3-D MFR, assuming the axial field ($B_Y$) is conserved
along the length of the MFR, as physically required, and assuming
$B_{\phi}\propto 1/a_{min}$ for the azimuthal field, where $a_{min}$
is the local minor radius of the MFR channel.  The accuracy of
this $B_{\phi}$ assumption is more uncertain, but it has the
effect of making the field overall more axial in the
legs of the MFR.  The MFR fields are naturally time-dependent.
The reference time that we use to define the initial field model,
corresponding to the apex maps in Figure~10, is when the leading
edge distance from the Sun, $R_{le}$, of the MFR is 1~au.  Even
though we do not even observe CME4 that far from the Sun, the
simple kinematic models in Figure~5 allow us to expand our model
CMEs out to 1~au (and beyond) if we want.  The advantage of a
1~au reference point is that this is where spacecraft have most
commonly measured CME fields in~situ, so it is easier to judge
what reasonable field values should be.  We are simply
assuming self-similar expansion for our MFRs, which means for
the field model that $B_Y$ and $B_{\phi}$ both scale as $1/R_{le}^2$.
With the VLA LOS encountering CME4 at a distance of about
9.4~R$_{\odot}$ (instead of 1~au), the actual fields the LOS will
be sampling will be roughly 524 times higher than in the 1~au
apex field maps shown in Figure~10.

     With the field model established it is now possible to
determine the MFR field along the VLA LOS for any time when the
LOS encounters CME4.  For each point along the LOS, it is
determined where the point is within the MFR, and this point
is then mapped to the apex where field magnitude can be
inferred using the scaling described above.  The rightmost
panel of Figure~10 shows the VLA LOS tracks through the
MFR channel (after mapping to the apex), displayed in 6-minute
time increments starting from right to left.  The Earth's
viewpoint would be from above the panel, and the CME apex
propagation direction would be to the right.

     Integration along the LOS of the field component parallel
to the LOS, multiplied by an assumed electron density ($n_e$),
yields an $RM$ prediction for that model.  For the field
model in Figure~10, with $B_t=+8$~nT and $B_p=+10$~nT, an
assumed constant internal CME density of $n_e=5000$~cm$^{-3}$
yields the predicted CME4 $RM$ signature shown in Figure~9(b).
This leads to an excellent fit to the observed positive-negative
MFR signature, but only after an arbitrary time shift of 45
minutes is added to the model prediction.  Our CME4
reconstruction apparently has the CME reaching the VLA LOS
about 45 minutes too early.  We believe this to be within
the expected uncertainties in the reconstruction, considering
how faint the leading edge is for this CME, and considering
that the reconstruction is a full-scale model of the CME shape
with no particular focus on reproducing its extent only
at the VLA LOS.

     Despite the 45 minute arrival time error, the model
accurately reproduces the $\sim 1$ hour encounter time
duration.  The positive-negative sign shift of $RM$ is
nicely reproduced, and this provides strong support for an
MFR structure for this CME, as noted above.  The assumed
$n_e=5000$~cm$^{-3}$ value at the VLA LOS encounter distance
would translate to an expected 1~au density of about
$n_e=9.5$~cm$^{-3}$, consistent with CME internal densities
typically observed at 1~au.

     It should be emphasized that the MFR field and density
model used to reproduce the $RM$ measurements will not be a
unique solution.  A trivial example of this would involve
increasing the assumed $B_t$ and $B_p$ values by some factor
and then decreasing $n_e$ by the same factor, which would yield
the exact same $RM$ signature.  A more significant alternate
model involves changing the overall polarity of the MFR.
The $B_t$ positive field model assumes the north leg of CME4
is the positive polarity leg.  If we instead assume the
northern leg is negative and $B_p$ is left-handed instead of
right-handed, this will also lead to a positive-negative $RM$
signature.  However, with the north leg of CME4 tilted
somewhat away from Earth, the positive $RM$ signature ends
up weaker than the following negative signature, making it
harder to fit the data.  This could be rectified simply by
assuming higher densities at the inner edge of the MFR.
But it should also be stated that we are not confident
that the assumed CME4 tilt angle ($\gamma_s$ in Table~1) is
really measured precisely enough to be sure that the north
leg is really tilted away from Earth.  Thus, we do not
claim a strong bias against the north leg being negative.

     Finally, we make no attempt to model the strong negative
$RM$ signature of the leg of CME4 (see Figure~9), as the
LOS is just grazing the leg.  Our 3-D reconstruction of
CME4 does not have the VLA LOS entering this leg, and
so clearly does not locate the leg correctly.  It is in
fact impossible for it to do so, since the functional form
used to define the shapes of the inner and outer edges of
the MFR are single-valued with angle \citep{bew09}.
In order for an LOS to encounter the inner edge of the CME
twice, it must by definition be double-valued, as in the
schematic picture in Figure~9(c).

\section{SDO/AIA Observes CME Remnant Plasma Falling Back to the Sun}

     We now focus our attention solely on CME2, the CME that is
disrupted by CME4 on 2024~August~16, at about UT~18:00.
While part of CME2 appears to be carried outward with CME4,
another part is replaced by a swarm of small downflows,
presumably channeling CME remnant material back down to the Sun.
We believe we can track this failed CME material all the way back
to the low corona, where it is observed in EUV images from
SDO/AIA \citep{jrl12}.

\begin{figure}[t]
\plotfiddle{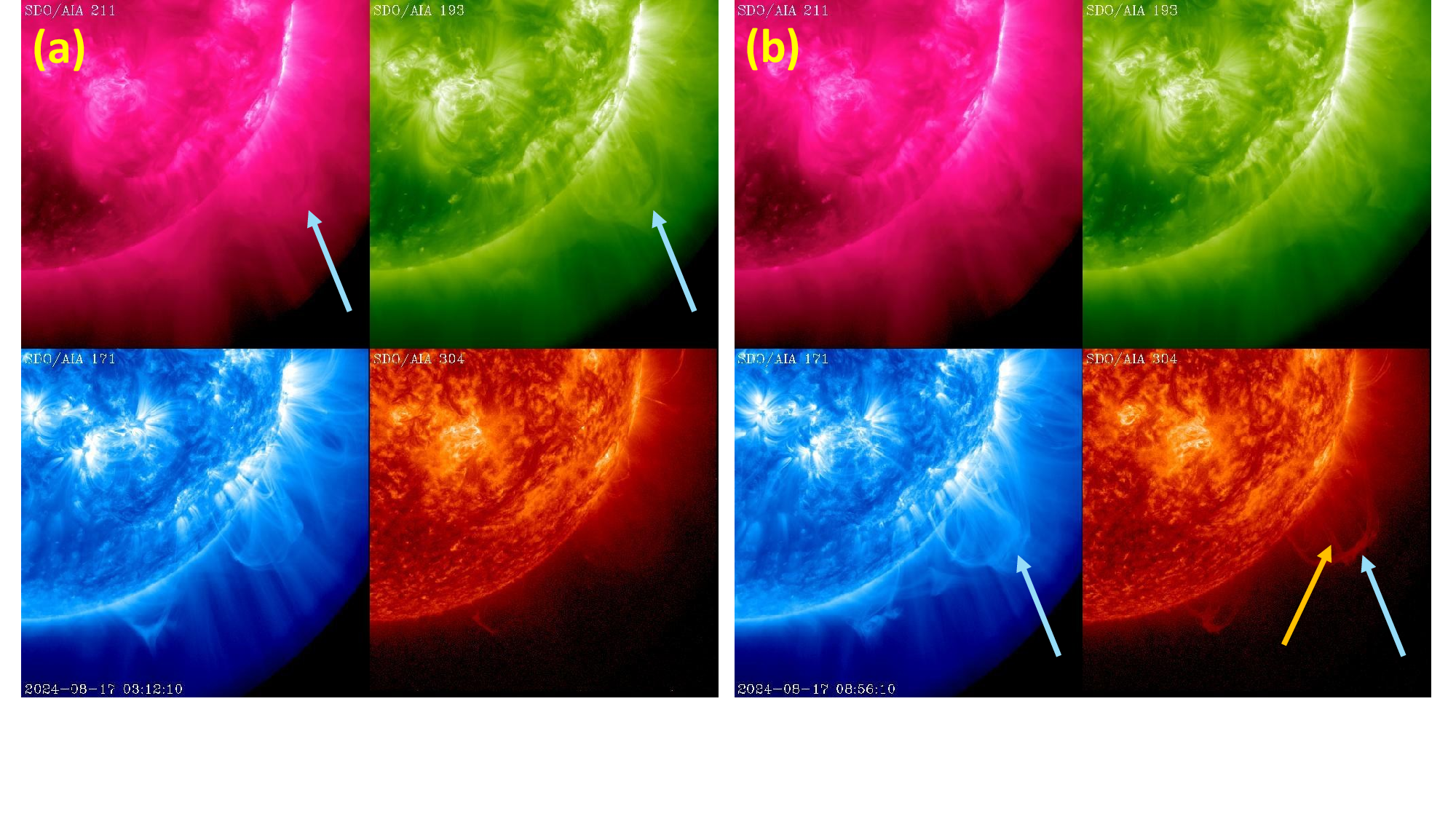}{2.7in}{0}{50}{50}{-240}{-50}
\caption{(a) Four SDO/AIA images from UT~03:12:10 on 2024~August~17,
  showing the solar corona in the 211~\AA, 193~\AA, 171~\AA, and 304~\AA\
  bandpasses, ordered from high temperature to low.  Light blue arrows
  identify a loop that brightens in the 211~\AA\ and 193~\AA\ bandpasses
  as material from the failed CME2 falls back into the low corona.
  (b) Four SDO/AIA images from UT~08:56:10, where the brightening loop
  is now seen in the lower temperature 171~\AA\ and 304~\AA\ bandpasses.
  An orange arrow in the 304~\AA\ frame identifies a coronal rain
  feature that flows down from the loop.  A movie version of this figure
  is available in the online version of this article covering
  two full days, August~16-17.}
\end{figure}
     Figure~11(a) and 11(b) show EUV images from UT~03:12:10 and
UT~08:56:10 on 2024~August~17, respectively, utilizing four SDO/AIA
bandpasses.  The bandpasses are shown in order of decreasing
temperature, with the characteristic temperatures being
$\log T=6.3$ for 211~\AA, $\log T=6.2$ for 193~\AA,
$\log T=5.8$ for 171~\AA, and $\log T=4.7$ for 304~\AA.
In movies of the first three of these bandpasses, above
the limb we perceive a faint downflow of material stretching
across much of the southwestern quadrant of the Sun, which we believe
corresponds to the remnants of CME2 falling back to the Sun.
This is most apparent above a particular arcade of coronal loops,
and light blue arrows in the Figure~11 identify a loop above this
arcade that first brightens at high temperatures (Figure~11a) and
later brightens at lower temperatures (Figure~11b).  In the lowest
temperature 304~\AA\ bandpass, the brightening loop appears very
much like a typical coronal rain condensation event
\citep[e.g.,][]{ss23}, which starts
with brightening at the top of the loop followed by numerous flows
downward from it to the surface of the Sun.  Other brightening
features associated with the CME2 downflow can also be seen in the
171~\AA\ and 304~\AA\ panels of Figure~11(b), but we will focus
on the region above the bright arcade.

\begin{figure}[t]
\plotfiddle{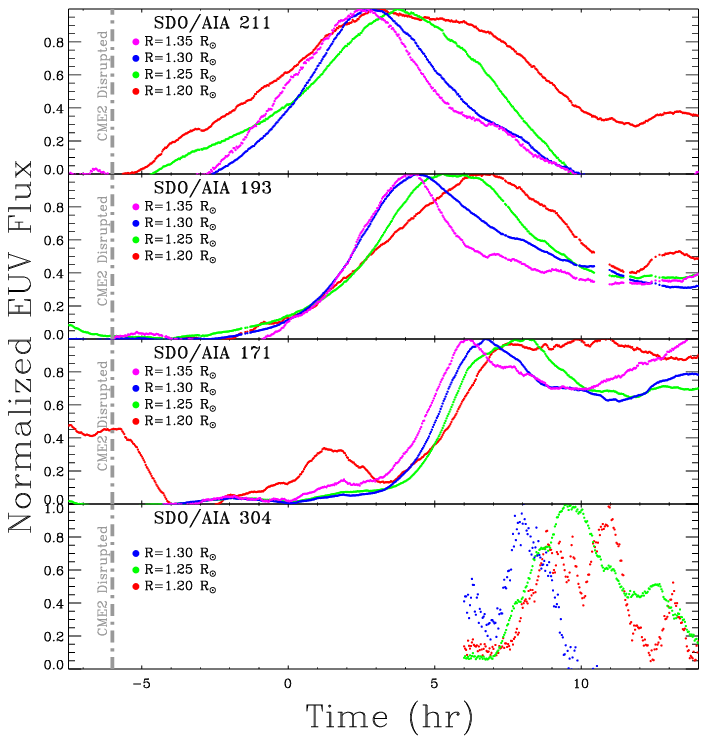}{3.4in}{0}{80}{80}{-240}{-300}
\caption{Normalized EUV brightnesses are measured for $12^{\circ}$ arcs
  above a loop arcade off the southwest limb of the Sun, and are
  plotted as a function of time, at four distances from the Sun,
  from $1.20-1.35$~R$_{\odot}$.  The measurements are for four
  SDO/AIA bandpasses (211~\AA, 193~\AA, 171~\AA, and 304~\AA),
  in order of decreasing temperature.
  The $t=0$ reference time is UT~0:00 on 2024~August~17, and a
  vertical dot-dashed line approximates the time when CME2 is
  disrupted.  Brightenings associated with downflowing material from
  this CME are seen in all bandpasses.  Evidence for downflow is
  particularly apparent in the 171~\AA\ panel, with the brightening
  clearly seen first at $R=1.35$~R$_{\odot}$ and then at successively
  lower heights.}
\end{figure}
     We measure EUV brightness in the four SDO/AIA bandpasses
across $12^{\circ}$ wide arcs above this arcade, at Sun-center
distances of 1.20~$R_{\odot}$, 1.25~$R_{\odot}$, 1.30~$R_{\odot}$,
and 1.35~$R_{\odot}$.  After subtracting a background value from
these brightnesses and then normalizing the peak flux to 1.0,
in Figure~12 we plot the resulting normalized brightnesses as
a function of time.  Clear brightening is seen for all bandpasses
and at all distances.  The evidence for this being a downflow is
strongest for the 171~\AA\ bandpass, with brightening first seen
at $R=1.35$~R$_{\odot}$ and then at successively lower heights.
With about an hour delay between the brightenings at
$R=1.35$~R$_{\odot}$ and at $R=1.20$~R$_{\odot}$, we estimate
a downflow speed of about $V=-30$ km~s$^{-1}$.  The is also seen
at 193~\AA, though there is little delay between $R=1.35$~R$_{\odot}$
and $R=1.30$~R$_{\odot}$.  The 211~\AA\ bandpass shows the expected
high-to-low brightening sequence for $R=1.35-1.25$~R$_{\odot}$,
but the lowest $R=1.20$~R$_{\odot}$ height seems anomalous.
Unrelated variations in the EUV background at $R=1.20$~R$_{\odot}$
appear to be responsible for this.  For the 304~\AA\
bandpass, the highest height of $R=1.35$~R$_{\odot}$ is not shown,
as brightening is not clearly apparent there.  The other
three heights seem to show the expected pattern of brightening
from high to low heights, but in this bandpass we associate this
more with the development of descending coronal rain features
than with the broad, faint downflow seen at higher temperatures.

\begin{figure}[t]
\plotfiddle{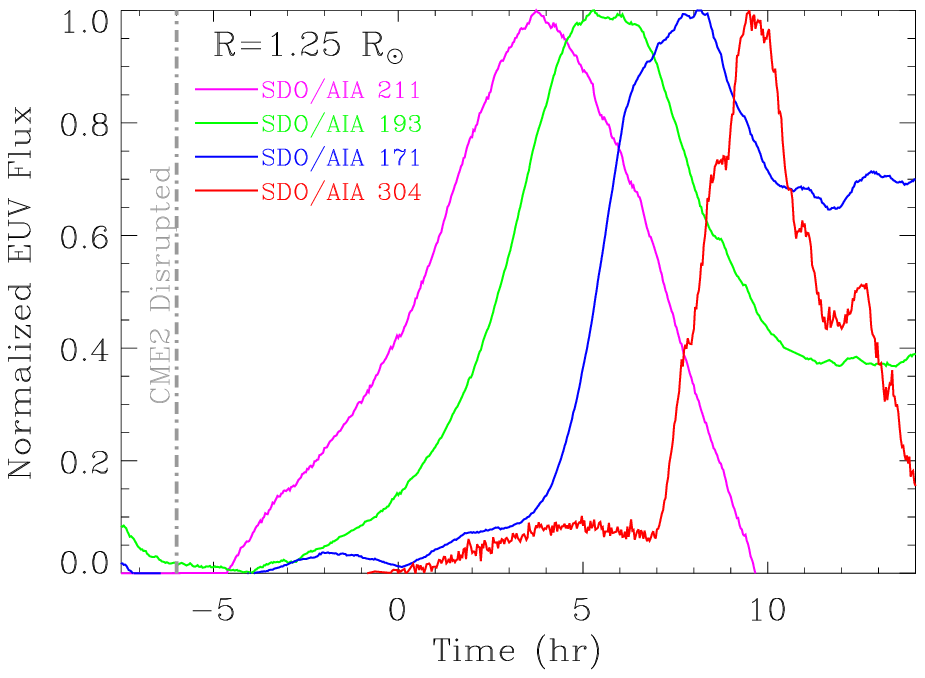}{3.0in}{0}{80}{80}{-260}{-310}
\caption{The $R=1.25$~R$_{\odot}$ EUV measurements from Figure~12
  are plotted for the four SDO/AIA bandpasses, illustrating the
  temperature dependence of the downflow signature, with the
  downflow first seen at the highest temperature bandpass
  (211~\AA) and then at successively lower temperatures, with
  the lowest temperature 304~\AA\ brightening occurring about
  eight hours after the 211~\AA\ brightening.}
\end{figure}
     The EUV light curves in Figure~12 not only demonstrate the
existence of a downflow, but they also show a clear temperature
dependence, with the downflow first seen in the highest
temperature bandpass, 211~\AA, and lastly in the lowest
temperature bandpass, 304~\AA.  Figure~13 illustrates
this temperature dependence more clearly, showing only the
$R=1.25$~R$_{\odot}$ light curves for the four bandpasses.
The brightness increase caused by the downflowing material
occurs about 8 hours later at 304~\AA\ compared with 211~\AA.
The most natural explanation for the
temperature dependence is that the falling CME2 remnant
material is
cooling with time.  It starts out at hot coronal temperatures
of $\log T\approx 6.3$, so the first downflowing material is
seen at those high temperatures.  Material that appears later,
possibly because it is falling from higher heights, has had
more time to radiatively cool, and is therefore observed at
lower temperatures.

     The interpretation of the 304~\AA\ data may be a bit
different.  Unlike at the higher coronal temperatures, where
movies of the SDO/AIA images suggest a faint, broad downflow
across much of the southwest quadrant, for 304~\AA\ the
impression is more of a number of coronal rain condensations
developing off the limb in this quadrant, which then flow
down to the solar surface.  Such condensations are occurring
before the CME downflow as well, though during the downflow
period their prevalence increases significantly, particularly
above the loop arcade being measured in Figures~12-13.  Rather
than the 304~\AA\ brightening being indicative of material that
is falling back to the Sun at these low $\log T\approx 4.7$
temperatures, it seems more likely that these are
condensations of the coronal material collected from the
downflows at higher temperatures, analogous to coronal rain
condensations that are also occurring before the CME downflow.

     Considerable effort has been made to model the thermal
instabilities that lead to coronal rain condensations
\citep{pa10,xf15,xl22,zl24,rk25,vl25}.
The observations of coronal rain presented here potentially
represent an opportunity to study this phenomenon in a
different context.  Rather than condensations of coronal
material heated in quasi-static coronal loops, these coronal
rain condensations instead seem to be triggered by flows of
coronal material from above.  It is worth noting once again
the prevalence of downflows seen by LASCO/C2 even
before CME2's remnants fall back to the Sun (see Figure~1).
If those smaller scale downflows are also progressing all the
way back to the low corona, is it possible that they also play
a role in stimulating coronal rain condensations?

\begin{figure}[t]
\plotfiddle{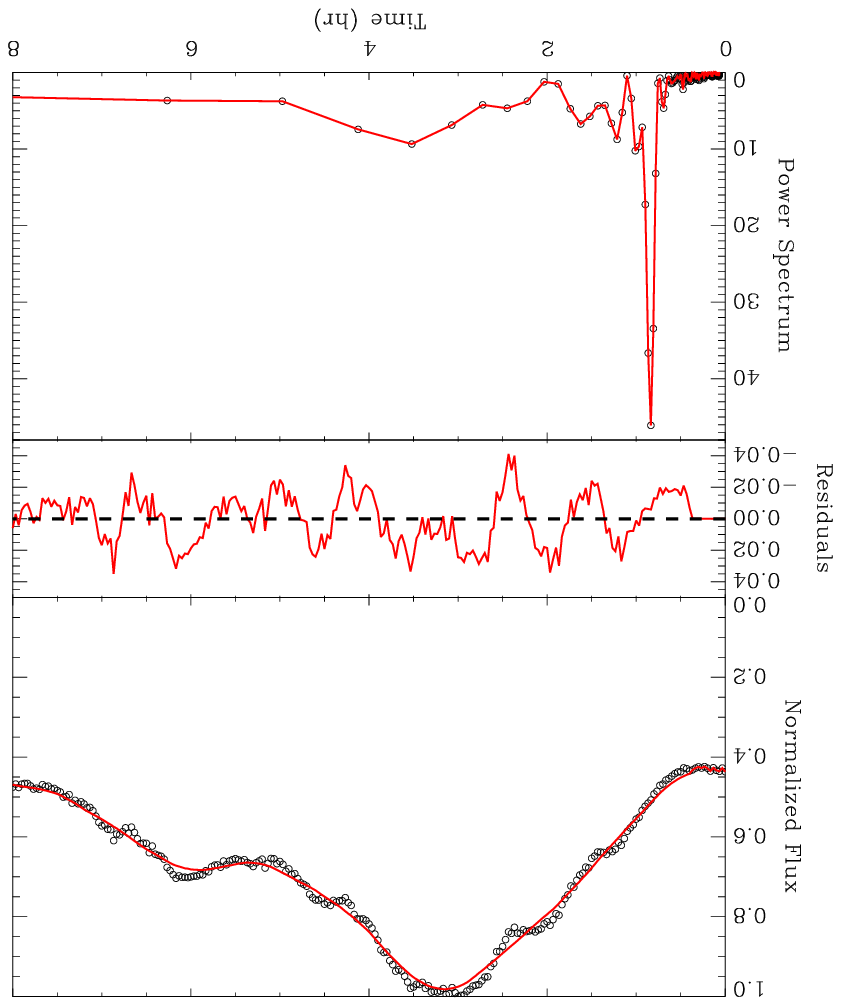}{3.4in}{180}{60}{60}{235}{310}
\caption{The top panel shows both the $R=1.25$~R$_{\odot}$ 304~\AA\
  light curve from Figure~13 and a heavily smoothed version of
  this curve.  Subtracting the latter from the former yields the
  middle panel, indicating a periodicity in the 304~\AA\ emission.
  The bottom panel is a Lomb-Scargle periodogram of this light
  curve, showing a strong peak at 50~min.}
\end{figure}
     There is one final characteristic of the 304~\AA\
measurements worth of note.  Close inspection of the 304~\AA\
light curve in Figure~13 reveals evidence of periodicity, on
top of the secular rise and fall of the 304~\AA\ emission at
$R=1.25$~R$_{\odot}$.  No such variation is observed for
the other EUV light curves in Figure~13, which are quite smooth.
The apparent 304~\AA\ periodicity is explored further in
Figure~14, where a version of the 304~\AA\ light curve smoothed
with a 25-bin running boxcar is subtracted from the unaltered
light curve.  The resulting residuals show a periodicity, which
we analyze using a Lomb-Scargle periodogram \citep{nrl76,jds82,jtv18}.
There is a very clear peak
at $t=50$~min.  Interpretation of this apparent periodicity
is difficult, as we find it impossible to visually distinguish
what it is that is causing it in the 304~\AA\ movies.  We do
note that oscillations of large solar prominences and filaments
are often found to have similar periods, with such oscillations
having a long history of study \citep{ia18}.
For example, in a survey of H$\alpha$ oscillations in filaments
by the Global Oscillation Network Group (GONG), \citet{ml18}
report a period distribution with a mean of $58\pm 15$ minutes.
However, we perceive no clear physical oscillation of the
brightening (and then fading) loop.

\section{Kinematic Modeling of the Downflowing CME Remnant Material}

     For the first time we believe we have detected material in
the low corona that is falling back to the Sun from very
high heights after a CME (CME2) is disrupted.  In the last section,
we estimated downflow velocities in the low corona of
about $V=-30$ km~s$^{-1}$.  For material falling ballistically
from very high heights, the expected speed of downflow when it
hits the surface of the Sun should be at or near the surface
escape speed of 618 km~s$^{-1}$. We are clearly nowhere near that.
Thus, the downflowing CME remnant material is not
falling unimpeded.  It is clearly experiencing a drag of some
sort.

     In this section, we present drag models of the downflow,
in order to provide plausible kinematic models for the falling
CME remnant plasma.  We use the same model as \citet{ymw02}
used for the core fallback events mentioned in Section~1.
This model assumes the drag force is proportional to the square of
the velocity difference between the falling material and the
ambient corona.  Mathematically, the radial motion of the
falling plasma is
\begin{equation}
\frac{d^2r}{dt^2}=-g_{\odot} \left(\frac{R_{\odot}}{r} \right)^2 -
  \frac{k(r)}{r} \left( \frac{dr}{dt} - V_b(r) \right)
  \left| \frac{dr}{dt} - V_b(r) \right|,
\end{equation}
where $g_{\odot}=274$ m~s$^{-2}$ is the solar gravitational constant
at the surface, $k(r)$ is a unitless drag coefficient, and $V_b(r)$
is the ambient solar wind flow velocity.  Unlike \citet{ymw02},
we assume a radial dependence for the drag coefficient, such that
\begin{equation}
k(r)=k_0 \left( \frac{r_0-R_{\odot}}{r-R_{\odot}} \right)^{\beta},
\end{equation}
where $k_0\equiv k(r_0)$, the drag coefficient at some reference
radius $r_0$.  We will use $r_0=1.25$~R$_{\odot}$ as our reference
radius, where our downflow velocity estimate of
$V=-30$ km~s$^{-1}$ applies.

     The goal of our modeling will be to determine what combination
of $k_0$ and $\beta$ values lead to downflows from heights of
$r=2.5-5.5$~R$_{\odot}$ that arrive at $r=1.25$~R$_{\odot}$ at the
right time and at the right velocity to match our SDO/AIA
observations.  Given our $V=-30$ km~s$^{-1}$ estimate, we will
consider velocities between $V=-15$ km~s$^{-1}$ and
$V=-45$ km~s$^{-1}$ to be acceptable.  As for time of arrival,
the time when CME2 is disrupted by CME4 is roughly UT~18:00
on August~16, which is $t=-6$~hr in Figures~12-13.
Based on Figure~13, the earliest EUV brightening in the 211~\AA\
bandpass is as soon as 2 hours after CME2's disappearance, though
the midpoint of the 211~\AA\ brightening (normalized flux
equals 0.5) is about 6 hours after, and the peak 211~\AA\ flux is
not reached until 10 hours after.  Based on the initial midpoint
increase of 211~\AA\ and the post-peak midpoint decrease of 304~\AA,
we estimate that the downflows are mostly observed over a
timespan of $6-17$~hr after CME2 disappears.  During this time,
material from the failed CME2 is presumed to be observed falling
from heights as far up as CME2's maximum leading edge height of
about 5.5~R$_{\odot}$.

     Finally, we need a model for the background solar wind speed,
$V_b(r)$.  For this, we use a prescription based on kinematic
measurements of streamer blobs \citep{ymw00}, which are
presumed to mirror the speed of the slow solar wind close to
the Sun.  The assumed solar wind speed (in km~s$^{-1}$) is
\begin{equation}
V_b(r) = \left\{ \begin{array} {c @{\quad\mbox{for}\quad} l}
  0 & r<2.4 R_{\odot} \\
  320\times \log \left( \frac{r}{2.4 R_{\odot}} \right) &
    2.4 R_{\odot} < r < 32 R_{\odot} \\
  360 & r>32 R_{\odot}.
  \end{array} \right.
\end{equation}
This is essentially a fit to the streamer blob
kinematic profile in Figure~9 from \citet{bew20a}.

\begin{figure}[t]
\plotfiddle{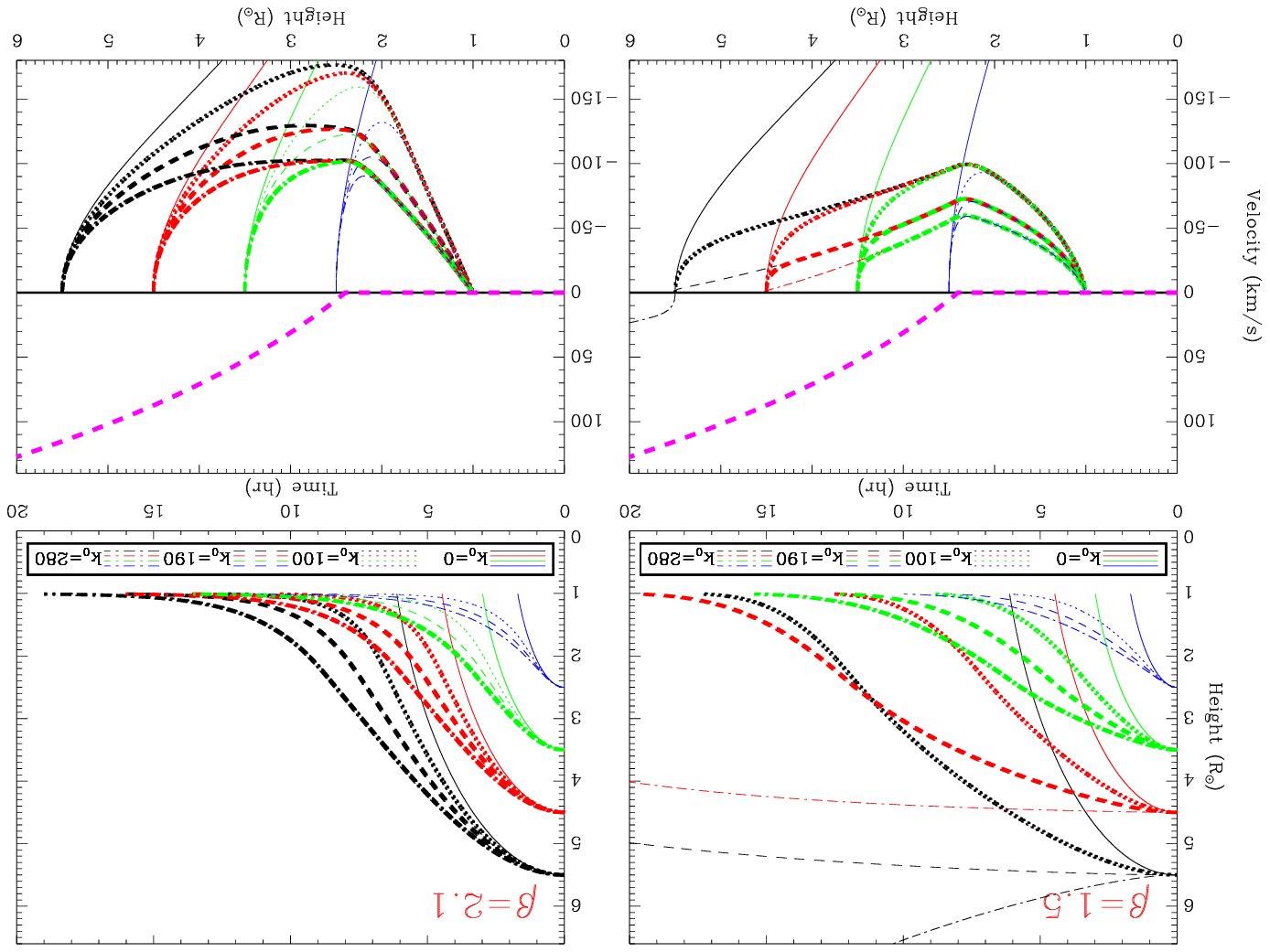}{3.45in}{180}{60}{60}{235}{315}
\caption{Kinematic drag models for downflowing CME2 remnant material.
  As depicted in the top panel, different line colors indicate
  different starting heights of 5.5~R$_{\odot}$ (black),
  4.5~R$_{\odot}$ (red), 3.5~R$_{\odot}$ (green), and
  2.5~R$_{\odot}$ (blue).  Different line styles indicate different
  drag coefficients at the reference radius of 1.25~R$_{\odot}$,
  in the range $k_0=0-280$.  The drag model assumes a height
  dependent drag coefficient with power law index $\beta$.
  The left (right) panels are for models with
  $\beta=1.5$ ($\beta=2.1$).  The lavender dashed lines in the
  bottom panels indicate the background solar wind model, $V_b(r)$,
  used in the drag model.  Thick lines indicate models
  that are deemed consistent with observation in terms of velocity
  and time of arrival at $r=1.25$~R$_{\odot}$.}
\end{figure}
     Figure~15 shows various kinematic models indicating how velocity
and height vary with time for material falling from four heights
ranging from $r=2.5-5.5$~R$_{\odot}$.  These trajectories are shown
for two values of $\beta$, $\beta=1.5$ and $\beta=2.1$, and five
values of the drag coefficient, in the range of $k_0=0-280$.
The $k_0=0$ profiles are purely ballistic trajectories.  Thick
lines in the figure indicate models that are deemed consistent
with SDO/AIA observations, with arrival at $r=1.25$~R$_{\odot}$ at
$t=6-17$~hr at a velocity in the $V=[-45,-15]$ km~s$^{-1}$ range.

     For $\beta=1.5$, it is $k_0=100$ that appears to work best,
with material falling from $r=3.5$, 4.5, and 5.5~R$_{\odot}$
arriving at $t=6.4$, 10.0, and 14.8~hr respectively; at a
velocity of $V=-39.7$ km~s$^{-1}$.  The model with
[$\beta$,$k_0$]=[1.5,100] therefore yields arrival times covering
the observed period of downflow reasonably well, and arriving in
the low corona at the correct velocity.  For $\beta=2.1$, it is
$k_0=280$ that works best, with material falling from
$r=3.5$, 4.5, and 5.5~R$_{\odot}$ arriving
at $t=6.9$, 9.5, and 12.3~hr respectively; at a velocity of
$V=-23.5$ km~s$^{-1}$.  There is clearly a best-fit parameter
correlation between $\beta$ and $k_0$, with low $k_0$ working
better for low $\beta$ and high $k_0$ working better for
high $\beta$.  One potential issue with the high beta models
is that they predict downflow speeds of $V\approx -100$ km~s$^{-1}$
or faster in the coronagraph field of view, compared with
the slower speed of $V=-38$ km~s$^{-1}$ estimated in Section~3.
Thus, although we would report $\beta=1.5-2.1$ and $k_0=100-280$
as being acceptable ranges for these parameters, we would
express some preference for the [$\beta$,$k_0$]=[1.5,100] model
at the low end of these ranges.

\section{Summary}

     We have presented the first example of observable EUV
downflows in the low corona originating from a CME observed to
be disrupted by a faster CME in coronagraphic observations of
the upper corona.  Our findings can be summarized as follows:
\begin{enumerate}
\item Based on coronagraphic imaging from SOHO/LASCO and STEREO,
  we provide full 3-D kinematic and morphological flux rope
  reconstructions of four CMEs that erupt at about the same
  time on 2024~August~16.  It is the second CME, CME2, that is
  the one that is disrupted by the fourth CME, CME4.
\item Part of CME2 is carried outward in the south leg of the
  CME4 flux rope.  This makes CME4 a very rare case where the
  leg of the flux rope is very well defined in the images,
  long after the CME leaves the coronagraph field of view.  The
  rest of CME2 disappears, and is replaced by a swarm of small
  jet-like downflows, presumed to be remnants of CME2 falling
  back to the Sun.
\item The four CMEs can be attributed to instabilities in the
  HCS, as all four have trajectories and orientations that
  mirror the HCS.
\item A field model is inserted into CME4 and used to
  reproduce radio $RM$ measurements made by the VLA of this
  CME.  A model with $B_t=+8$~nT and $B_p=+10$~nT (at 1~au)
  leads to a good fit to the data, assuming $n_e=5000$~cm$^{-3}$
  along the LOS within the CME.  The positive-negative sign
  shift observed for the $RM$ values provides strong support
  for the MFR paradigm of CME structure, as the azimuthal
  field of the MFR around the central axis naturally yields
  this sign change.
\item Roughly 6-17~hr after CME2 is disrupted, downflows are
  observed off the limb across the southwest quadrant of the
  Sun by SDO/AIA in the 211~\AA, 193~\AA, and 171~\AA\
  bandpasses, with an estimated velocity of $V=-30$ km~s$^{-1}$.
  There is a clear temperature dependence, with the downflows
  first seen in the highest temperature 211~\AA\ bandpass,
  followed successively by 193~\AA\ and 171~\AA.  This may
  due to radiative cooling of the downflowing plasma.
  A response at 304~\AA\ is also observed, with condensations
  observed that develop into coronal rain.  A coronal rain
  response to the CME2 remnant downflow suggests that perhaps
  smaller scale downflows that are often seen in coronagraphic
  images \citep[e.g.,][]{ymw99} may contribute to coronal
  rain activity more generally.
\item Periodic intensity variations are observed in the 304~\AA\
  observations of a particular loop that brightens in response
  to the CME downflow.  The cause of the 50~min periodicity is
  uncertain.
\item The observed speed of descent is much too slow for this
  to be a simple ballistic downflow, so the flow must be
  impeded.  We provide estimates of the full kinematic profile of
  the downflowing CME material using kinematic drag models
  constructed assuming an aerodynamic-like drag proportional
  to the square of the velocity difference with the ambient
  corona.
\end{enumerate}

\acknowledgments

We acknowledge financial support from the Office of Naval
Research.  This article involves the use of data products
provided by the NSO Integrated Synoptic Program (NISP), with data
acquired by SOLIS instruments operated by NISP/NSO/AURA/NSF
(see https://nso.edu/data/nisp-data).  The Karl G.\ Jansky
Very Large Array is an instrument of the National Radio Astronomy
Observatory.  The National Radio Astronomy Observatory and
Green Bank Observatory are facilities of the U. S. National
Science Foundation operated under cooperative agreement by
Associated Universities, Inc.


\begin{thebibliography}{}

\bibitem[Antolin et al.(2010)]{pa10}
Antolin, P., Shibata, K., \& Vissers, G. 2010, ApJ, 716, 154
\bibitem[Arregui et al.(2018)]{ia18}
Arregui, I., Oliver, R., \& Ballester, J. L. 2018, LRSP, 15, 3
\bibitem[Billings(1966)]{deb66}
Billings, D. E. 1966. A Guide to the Solar Corona (New York: Academic
  Press)
\bibitem[Bird et al.(1985)]{mkb85}
Bird, M. K., Volland, H., Howard, R. A., et al. 1985, Sol.~Phys., 98, 341
\bibitem[Bothmer \& Schwenn(1998)]{vb98}
Bothmer, V., \& Schwenn, R. 1998, Ann.~Geophys., 16, 1
\bibitem[Brueckner et al.(1995)]{geb95}
Brueckner, G. E., Howard, R. A., Koomen, M. J., et al. 1995, Sol.~Phys.,
  162, 357
\bibitem[Chen et al.(1997)]{jc97}
Chen, J., Howard, R. A., Brueckner, G. E., et al. 1997, ApJ, 490, L191
\bibitem[Fang et al.(2015)]{xf15}
Fang, X., Xia, C., Keppens, R., \& Van Doorsselaere, T. 2015, ApJ, 807, 142
\bibitem[Harra et al.(2016)]{lkh16}
Harra, L. K., Schrijver, C. J., Janvier, M., et al. 2016, Sol. Phys., 291, 1761
\bibitem[Hess \& Wang(2017)]{ph17}
Hess, P., \& Wang, Y. -M. 2017, ApJ, 850, 6
\bibitem[Howard et al.(2008)]{rah08}
Howard, R. A., Moses, J. D., Vourlidas, A., et al. 2008, Space Sci. Rev.,
  136, 67
\bibitem[Jensen et al.(2018)]{eaj18}
Jensen, E. A., Heiles, C., Wexler, D., et al. 2018, ApJ, 861, 118
\bibitem[Jensen et al.(2025)]{eaj25}
Jensen, E. A., Manchester, W. B., Kooi, J. E., et al. 2025, ApJ, 987, 156
\bibitem[Joshi et al.(2013)]{ncj13}
Joshi, N. C., Srivastava, A. K., Filippov, B., et al. 2013, ApJ, 771, 65
\bibitem[Kay et al.(2013)]{ck13}
Kay, C., Opher, M., \& Evans, R. M. 2013, ApJ, 775, 5
\bibitem[Kazachenko(2023)]{mdk23}
Kazachenko, M. D. 2023, ApJ, 258, 104
\bibitem[Keppens et al.(2025)]{rk25}
Keppens, R., Zhou, Y., \& Xia, C. 2025, LRSP, 22, 4
\bibitem[Kooi et al.(2017)]{jek17}
Kooi, J. E., Fischer, P. D., Buffo, J. J., \& Spangler, S. R. 2017,
  Sol. Phys., 292, 56
\bibitem[Kooi et al.(2022)]{jek22}
Kooi, J. E., Wexler, D. B., Jensen, E. A., et al. 2022, FrASS, 9, 841866
\bibitem[Lemen et al.(2012)]{jrl12}
Lemen, J. R., Title, A. M., Akin, D. J., et al. 2012, Sol.~Phys., 275, 17
\bibitem[Lepping et al.(1990)]{rpl90}
Lepping, R. P., Jones, J. A., \& Burlaga, L. F. 1990, JGR, 95,
  11957
\bibitem[Li et al.(2021)]{tl21}
Li, T., Chen, A., Hou, Y., et al. 2021, ApJ, 917, L29
\bibitem[Li et al.(2022)]{xl22}
Li, X., Keppens, R., \& Zhou, Y. 2022, ApJ, 926, 216
\bibitem[Liakh \& Jenkins(2025)]{vl25}
Liakh, V., \& Jenkins, J. 2025, Sol.~Phys., 300, 147
\bibitem[Liewer et al.(2009)]{pcl09}
Liewer, P. C., de Jong, E. M., Hall, J. R., et al. 2009, Sol.~Phys., 256, 57
\bibitem[Lomb(1976)]{nrl76}
Lomb, N. R. 1976, Ap\&SS, 39, 447
\bibitem[Lu et al.(2024)]{zl24}
Lu, Z., Chen, F., Guo, J. H., et al. 2024, ApJ, 973, L1
\bibitem[Lugaz et al.(2009)]{nl09}
Lugaz, N., Vourlidas, A., \& Roussev, I. I. 2009, Ann.~Geophys., 27, 3479
\bibitem[Luna et al.(2018)]{ml18}
Luna, M., Karpen, J., Ballester, J. L., et al. 2018, ApJS, 236, 35
\bibitem[Lynch(2020)]{bjl20}
Lynch, B. J. 2020, ApJ, 905, 139
\bibitem[Nieves-Chinchilla et al.(2018)]{tnc18}
Nieves-Chinchilla, T., Linton, M. G., Hidalgo, M. A., \& Vourlidas, A.
  2018, ApJ, 861, 139
\bibitem[\c{S}ahin et al.(2023)]{ss23}
\c{S}ahin, S., Antolin, P., Froment, C., \& Schad, T. A. 2023, ApJ, 950, 171
\bibitem[Sanchez-Diaz et al.(2017)]{esd17}
Sanchez-Diaz, E., Rouillard, A. P., Davies, J. A., et al. 2017, ApJ, 835, L7
\bibitem[Scargle(1982)]{jds82}
Scargle, J. D. 1982, ApJ, 263, 835
\bibitem[Sheeley et al.(2009)]{nrs09}
Sheeley, N. R., Jr., Lee, D. D. -H., Casto, K. P., Wang, Y. -M., \& Rich, N. B.
  2009, ApJ, 694, 1471
\bibitem[Sheeley \& Wang(2014)]{nrs14}
Sheeley, N. R., Jr., \& Wang, Y. -M. 2014, ApJ, 797, 10
\bibitem[Sheeley \& Wang(2007)]{nrs07}
Sheeley, N. R., Jr., \& Wang, Y. -M. 2007, ApJ, 655, 1142
\bibitem[Sheeley \& Wang(2002)]{nrs02}
Sheeley, N. R., Jr., \& Wang, Y. -M. 2002, ApJ, 579, 874
\bibitem[Thernisien et al.(2006)]{afrt06}
Thernisien, A. F. R., Howard, R. A., \& Vourlidas, A. 2006, ApJ, 652, 763
\bibitem[VanderPlas(2018)]{jtv18}
Vanderplas, J. T. 2018, ApJS, 236, 16
\bibitem[Vourlidas et al.(2013)]{av13}
Vourlidas, A., Lynch, B. J., Howard, R. A., \& Li, Y. 2013, Sol.~Phys.,
  284, 179
\bibitem[Vourlidas et al.(2025)]{av25}
Vourlidas, A., Paouris, E., Linton, M. G., et al. 2025, ApJ, 995, L38
\bibitem[Wang \& Hess(2018)]{ymw18}
Wang, Y. -M., \& Hess, P. 2018, ApJ, 859, 135
\bibitem[Wang \& Sheeley(2002)]{ymw02}
Wang, Y. -M., \& Sheeley, N. R., Jr. 2002, ApJ, 567, 1211
\bibitem[Wang et al.(1999)]{ymw99}
Wang, Y. -M., Sheeley, N. R., Jr., Howard, R. A., St. Cyr, O. C., \&
  Simnett, G. M. 1999, GRL, 26, 1203
\bibitem[Wang et al.(2000)]{ymw00}
Wang, Y. -M., Sheeley, N. R., Jr., Socker, D. G., Howard, R. A., \& Rich, N. B.
  2000, JGR, 105, 25133
\bibitem[Wang et al.(1998)]{ymw98}
Wang, Y. -M., Sheeley, N. R., Jr., Walters, J. H., et al. 1998, ApJ, 498, L165
\bibitem[Wood \& Hess(2025)]{bew25}
Wood, B. E., \& Hess, P. 2025, ApJ, 980, 113
\bibitem[Wood et al.(2020a)]{bew20a}
Wood, B. E., Hess, P., Howard, R. A., Stenborg, G., \& Wang, Y. -M. 2020a,
  ApJS, 246, 28
\bibitem[Wood \& Howard(2009)]{bew09}
Wood, B. E., \& Howard, R. A. 2009, ApJ, 702, 901
\bibitem[Wood et al.(2016)]{bew16}
Wood, B. E., Howard, R. A., \& Linton, M. G. 2016, ApJ, 816, 67
\bibitem[Wood et al.(2020b)]{bew20b}
Wood, B. E., Tun-Beltran, S., Kooi, J. E., Polisensky, E. J., \&
  Nieves-Chinchilla, T. 2020b, ApJ, 896, 99
\bibitem[Wood et al.(2017)]{bew17}
Wood, B. E., Wu, C. -C., Lepping, R. P., et al. 2017, ApJS, 229, 29
\bibitem[Yashiro et al.(2004)]{sy04}
Yashiro, S., Gopalswamy, N., Michalek, G., et al. 2004, JGR, 109, A07105

\end{thebibliography}
\end{document}